%% file: main.tex
\documentclass[%
 reprint, 
 amsmath,amssymb,
 aps, 
prx,
superscriptaddress,
english
]{revtex4-2}

\usepackage{babel}

\usepackage{graphicx}
\usepackage{tikz}

\begin{document}

\preprint{APS/123-QED}

\title{A Constant Measurement Quantum Algorithm for Graph Connectivity}%

\author{Maximilian Balthasar Mansky}
\email{Contact author: maximilian-balthasar.mansky@ifi.lmu.de}
\affiliation{Department of Informatics, LMU Munich, 80538 Munich, Germany}
\author{Chon-Fai Kam}
\affiliation{Department of Physics, SUNY Buffalo, New York 14260, USA}
\author{Claudia Linnhoff-Popien}
\affiliation{Department of Informatics, LMU Munich, 80538 Munich, Germany}

\date{\today}

\begin{abstract}

We introduce a novel quantum algorithm for determining graph connectedness using a constant number of measurements. The algorithm can be extended to find connected components with a linear number of measurements. It relies on non-unitary abelian gates taken from ZX calculus. Due to the fusion rule, the two-qubit gates correspond to a large single action on the qubits. The algorithm is general and can handle any undirected graph, including those with repeated edges and self-loops. The depth of the algorithm is variable, depending on the graph, and we derive upper and lower bounds. The algorithm exhibits a state decay that can be remedied with ancilla qubits. We provide a numerical simulation of the algorithm.
\end{abstract}

\maketitle


\section{Introduction}


The emergence of quantum computing has brought forth new paradigms for addressing computational problems \cite{nielsen_quantum_2010}. Well-designed quantum algorithms can offer significant speedups over their classical counterparts, with potential advantages stemming from quantum resources \cite{coecke2016mathematical, schmid2020type} such as non-classicality \cite{tilma2010entanglement,catani2018non, kam2023coherent}, multipartite entanglement \cite{horodecki2009quantum, bengtsson2017geometry, schmid2023understanding} and quantum non-locality \cite{scarani2019bell, wolfe2020quantifying, rosset2020type}. To assess the advantage of one quantum algorithm over another, one needs the fundamental concept of query complexity, which quantifies how many times an algorithm needs to access or query the input data to solve a problem \cite{nielsen_quantum_2010, montanaro_quantum_2016}. 

The scaling of an algorithm is an important metric on its feasibility for application \cite{cormen_introduction_2009}. While all problems can be solved by brute force, iterating over all possible values, the often exponential scaling of the problem space makes this approach time-consuming. Algorithm runtimes are generally classed into groups with respect to their scaling of necessary number of steps versus an increase in problem size. Broadly these are \textsf{NP}, non-deterministic polynomial with exponential runtime, \textsf{P} for polynomial, linear runtime and constant time. The big-O notation $\mathcal{O}(f(n))$ gives the asymptotic behavior for the runtime of the algorithm as a function of input size \cite{sipser_introduction_1996}. 

Algorithms of constant time perform at a speed independent of the size of the input, whereas for linear algorithms the time necessary to compute the outcome of the algorithm scales linearly. Similarly, for polynomial algorithms the runtime can be expressed as a polynomial function of the input. For the asymptotic, only the largest factor in that polynomial matters.  Naturally, exponential algorithms can only run for small problem sizes, as the cost of computation quickly exceeds the available resources. 

In the context of quantum computing and its algorithms, there exist additional complexity classes that go beyond classical ones. A key class relevant to this work is \textsf{BQP} (bounded-error quantum polynomial time), which is defined as the class of decision problems that can be solved by a quantum computer in polynomial time and with a probability of error that is bounded \cite{bernstein_quantum_1997, nielsen_quantum_2010}. A decision problem involves providing a binary output (either \emph{yes} or \emph{no}) based on a given input. For a problem to be in \textsf{BQP}, the quantum computer must return the correct binary answer with a probability of at least $2/3$. This means that the quantum algorithm is expected to produce the correct result most of the time, with a small probability of error.

Quantum computing excels in solving problems where relationships and patterns among entities must be explored simultaneously—a property well-suited to graph structures. Graph problems, such as connectivity, benefit from quantum features like superposition, which allows the examination of multiple pathways concurrently. Graph theory has been a central topic in computer science for decades, as it provides a natural abstract framework for modeling relationships between objects in various real-world problems, from social networks \cite{ediger_massive_2010, traud_social_2012} to biological systems \cite{stam_graph_2007, pavlopoulos_using_2011, farahani_application_2019}, communication networks \cite{monge_theories_2003}, and even quantum physics \cite{hein_entanglement_2006, coecke_compositional_2010, de_beaudrap_quantum_2014, gnatenko_geometric_2022, vesperini_entanglement_2024} itself. The problem of graph connectivity, which involves determining whether all the nodes in a graph can be reached by traversing its edges, is a fundamental problem in graph theory \cite{gross_handbook_2004}. It acts as fundamental building block like network reliability \cite{ball_network_1995}, clustering \cite{schaeffer_graph_2007} and routing \cite{erciyes_distributed_2013}. 

The graph connectivity problem can be solved classically in linear time by using a breadth-first approach. Faster classical approaches exist by parallelizing the computation \cite{shiloach_ologn_1982}. It can also be reformulated as a search problem with a quadratic runtime by looking for the minimum spanning tree within the graph \cite{gross_handbook_2004}. The search operation within the algorithm can be done with a quantum computer to construct a hybrid algorithm. This approach has been taken for quantum algorithms for connectivity (which is where we improve upon), strong connectivity, minimum spanning tree, and single source shortest paths \cite{durr_quantum_2006}. These algorithms are hybrid in nature, each applying Grover's quantum search algorithm \cite{grover_fast_1996} to accelerate part of the existing classical algorithms and to efficiently query a matrix and array model of the problem. 


In this work, we focus on the problem of graph connectivity as well as graph fragments for undirected graphs. The requirements of the algorithms are discussed first, indicating how one can move from global structural requirements of the decision problem to individual requirements for the gates used in the construction of the algorithm. The background section \ref{sec:background} contains the knowledge necessary to understand the construction. Section \ref{sec:algorithm} explicitly shows the construction of a quantum circuit that can determine the connectivity and identify the connected components of a graph in with a constant amount of measurements. It relies on the contraction rules of non-unitary gates taken from ZX calculus. The depth of the quantum circuit depends on the graph and we provide upper and lower bounds for the depth, together with an approximate average case. The algorithm exhibits state decay, which can be remedied using ancilla qubits. The discussion closes our paper.

\section{Requirements for the algorithm}\label{Requirements}

Before constructing the algorithm, it is advisable to consider what requirements have to be placed on it. The connectedness of a graph is a global property, decided only by the connections created by the edges of a graph. A disconnected graph decomposes into two or more connected components, which should be distinguishable in some way. The individual edges are generally indistinguishable, meaning that no edge should be special compared to the others.

The constraints of the graph structure are instructive to the requirements on the individual gates needed for implementing the edges. The entangling operations have to be abelian, as the list of edges is generally unsorted. Sorting the list classically requires at least as much time as finding connected components (The runtime of  a sorting algorithm on the edges is $\mathcal{O}(m \log m)$). This requirement rules out the obvious alternative to the present approach, creating entanglement using CNOT gates. All edges have to be of the same type, as implementing specific gates for particular edges would again require some insight into the graph structure. The gate representing an edge also needs to be nilpotent, $N^2 = N$. Otherwise, loops or repeated edges will cause cancellations that destroy the information of connectedness.

We are allowed to pick at most one node to be special, for example as a starting point to a connected fragment. The qubit corresponding to that node then can receive additional gates. In our implementation of the algorithm, this is not necessary.

\section{Background}\label{sec:background}

In this section, we provide the essential background needed to present our algorithm, both from the perspective of graph theory and the ZX calculus.

\subsection{Graph structure}\label{ssec:graph-structure}

A graph $\mathcal{G} = (N, E)$ consists of nodes $N$ and edges $E=\{n_i, n_j\}$, where the edges are undirected, meaning $\{n_i, n_j\} = \{n_j, n_i\}$. We denote the number of nodes in $\mathcal{G}$ as $n = |N|$ and the number of edges as $m = |E|$. $\mathcal{G}$ is said to be connected if, for every pair of distinct nodes $n_i$, $n_j$ in $N$, there exists a path of edges in $E$ that connects $n_i$ to $n_j$. In other words, there is a sequence of edges that allows traversal from any node to any other node in $\mathcal{G}$. A connected component is a subgraph $\mathcal{G}_k$ of $\mathcal{G}$ that is internally connected, but has no edges linking it to other components, meaning there are no edges between the nodes of $\mathcal{G}_k$ and those of any other component $\mathcal{G}_l$. For the purposes of the algorithm, we assume that the edges are provided as an unsorted list of node tuples. 

\subsection{ZX-calculus}

\begin{figure}[h!]
	$\vcenter{\hbox{\includegraphics{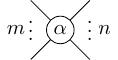}}}%
        := |0\rangle^{\otimes m} \langle 0 |^{\otimes n} + e^{i\alpha} |1\rangle^{\otimes m}\langle 1|^{\otimes n}$\\
	$\vcenter{\hbox{\includegraphics{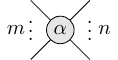}}}
        := |-\rangle^{\otimes m} \langle - |^{\otimes n} + e^{i\alpha} |+\rangle^{\otimes m}\langle +|^{\otimes n}$
        \caption{The main elements of ZX calculus are the spiders. Contrary to standard quantum circuit elements, the number of inputs and outputs does not have to be fixed.}
        \label{fig:spider-def}
        \end{figure}

The ZX-calculus is a graphical language to describe quantum circuits and, more generally, linear maps~\cite{coecke_interacting_2008,coecke_picturing_2017}. For an introduction coming from a quantum computing background, we refer to~\cite{van_de_wetering_zx-calculus_2020}, for the perspective from its category theory roots, we refer to~\cite{coecke_interacting_2011}.
A ZX-diagram is made of spiders representing multi-dimensional tensors in either $Z$ or $X$ basis which are connected via wires. The definition is given in figure \ref{fig:spider-def}.

The Z and X spiders are commonly color coded as green/white and red/gray across the literature. A third common object is the Hadamard operator, generally presented as a yellow box or a dashed wire. The algorithm presented here makes use only of Z spiders, which we will present as white circles throughout. 

\begin{figure}[h!]
    \centering
	\includegraphics{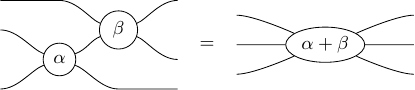}
        
    \caption{The  ingredients of ZX calculus  relevant to this work. The contraction rule allows to combine two spiders of the same color to one larger spider \cite{van_de_wetering_zx-calculus_2020}. The other rules of ZX calculus play no rule in the construction of the algorithm.}
    \label{fig:zx} 
\end{figure}

\begin{figure}[hb]
    \centering
	\includegraphics{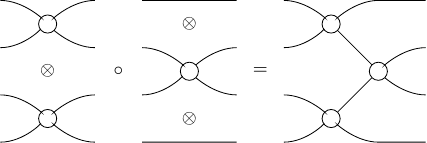}
    \caption{Composition in ZX diagrams. Similar to the standard diagrammatic framework of quantum circuit drawings, parallel elements are composed using the tensor product and sequential operations are concatenated using the matrix multiplication.}
    \label{fig:zx-composition}
\end{figure}

ZX-diagrams can be composed either via the tensor product by placing them side by side, or via matrix multiplication, by joining wires. This is illustrated in figure~\ref{fig:zx-composition}. There exist complete rule sets to transform ZX-diagrams into equivalent ones~\cite{vilmart-near-optimal-2018}, for instance, we can fuse adjacent spiders of the same color into a single one, shown in figure~\ref{fig:zx}. In fact, we only need this fusion rule to construct the quantum algorithm.


Individual spiders do not correspond to unitary operations, except single wire input and output. Indeed, ZX calculus can be used to generally reason about linear maps, rather than the more restricted special unitary group. In principle, ZX calculus can be used to represent any linear map and therefore every quantum circuit \cite{wang_representing_2023}.


Composition works in the same way as the standard quantum circuit formulation \cite{nielsen_quantum_2010}. This is illustrated in figure \ref{fig:zx-composition}. For parallel operations, the corresponding matrices are combined using the tensor product, sequential operations are concatenated using the matrix multiplication \cite{van_de_wetering_zx-calculus_2020}. 

\subsection{Non-unitary gates}

The Z and X spiders individually are non-unitary operations. Non-unitary gates are universal for computation, however their application is inherently probabilistic \cite{terashima_nonunitary_2005}. The probability of success for the gate depends on the input states. For a general state $|\psi\rangle$ the probability of success for a non-unitary gate will less than one, leading to an exponential runtime for non-unitary calculations. We will show that for some states and gates such as those used in the presented algorithm, the success probability is always one.

Non-unitary gates have a complex-valued $2^n \times 2^n$ matrix representation in the same way as unitary gates. The quantum state after application of a non-unitary operation $N$ is 
\begin{equation}
    \frac{N |\psi\rangle}{\sqrt{\langle \psi| N^\dagger N |\psi\rangle}}
\end{equation}
where $N$ is a non-unitary transformation, in the present case the white spider of figure \ref{fig:spider-def}. The probability of success upon measurement is given as \cite{terashima_nonunitary_2005}:

\begin{equation}
    p(|\psi\rangle; c) = |c|^2\langle \psi | N^\dagger N | \psi\rangle
\end{equation}

with a normalization constant $c$. The success probability relates to the eigenvalues of $N$. From definition of a spider given in figure \ref{fig:spider-def}, it is apparent that it acts as a projector. For a given state $|\psi\rangle$, the Z spider has success probabilities

\begin{equation}
    \text{Z spider}(\alpha) |\psi\rangle = \begin{cases}
        1 & a|0\ldots 0\rangle + be^{i\alpha}|1\ldots 1\rangle\\
        0 & \text{otherwise}
    \end{cases}
\end{equation}

This means that for non GHZ-like state, the operation surely fails, For the parameterized $|\alpha\text{GHZ}\rangle$ state,

\begin{equation}
    |\alpha\text{GHZ}\rangle = |0\ldots 0 \rangle + e^{i\alpha} |1\ldots 1\rangle
\end{equation} 

the Z spiders have a success probability of $1$. Effectively, the spider acts as a projector within the reduced state space 
\begin{equation}
    \operatorname{span}\{|0\ldots 0\rangle, |1\ldots 1\rangle\}
\end{equation}

This state space is sufficient for the presented algorithm. The spiders are expressible as a matrix. For two wires, the Z spider with parameter $\alpha$ is given as
\begin{equation}
Z(\alpha)
    = \begin{pmatrix}
        1 & 0 & 0 & 0\\
        0 & 0 & 0 & 0\\
        0 & 0 & 0 & 0\\
        0 & 0 & 0 & e^{i\alpha}
    \end{pmatrix}
\end{equation}

\begin{figure}[h!]
	$\vcenter{\hbox{\includegraphics{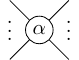}}}
    = \begin{pmatrix}
        1 & 0 & \ldots &  0 & 0\\
        0 & 0 &  &  & 0\\
        \vdots & & \ddots & & \vdots\\
        0 &  &  & 0 & 0\\
        0 & 0 & \ldots & 0 & e^{i\alpha}
    \end{pmatrix}$
    \caption{The general matrix representation of a multi-legged Z spider. Only the first and last entry of the matrix have an entry, $1$ and $exp(i\alpha$ respectively.}\label{fig:large-spider-matrix}
\end{figure}

Larger spiders follow the same pattern, with a matrix constructed with $1$ in the first entry, $\exp(i\alpha)$ in the last one, as shown in figure \ref{fig:large-spider-matrix}.

\section{Graph connectedness algorithm}\label{sec:algorithm}

We introduce the construction of the quantum graph connectedness algorithm and analyze its runtime and depth in comparison to classical algorithms. 


\subsection{Construction}\label{ssec:construction}

In the construction of our algorithm, each node $n_i$ of the graph $\mathcal{G}$ is mapped to a qubit $q_i$. The edges $m_i = \{n_i, n_j\}$ are mapped to two-qubit gates between the respective qubits $\{q_i, q_j\}$. The qubits are entangled via the edges to a large multi-qubit entangled state that can be measured to detect the connectedness between the qubits. Each two-qubit gate is one $Z$ spider explained in the preceding section. The two-qubit spiders are mathematically equivalent to large spiders that connect all qubits corresponding to a connected component. Its action creates a large GHZ state for each connected component.



The non-unitary spiders of ZX calculus fulfil this requirement. They act as projectors from the ground state to a parameterized GHZ state and they retain the desired abelian property. The fact that they are contractable to a single spider means that all qubits within a given graph fragment $\mathcal{G}_k$ are entangled in the same way. Conversely, for calculating the final state, one only needs to consider one spider corresponding to a connected component.

\begin{figure*}
    \centering
	\includegraphics{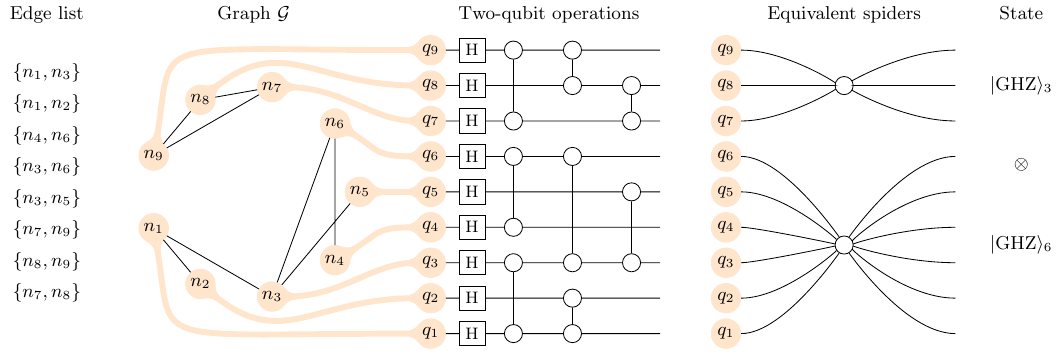}
        
    \caption{Illustration of the graph component algorithm for a two-piece graph. It shows from left to right the transformation from a list of edges and its associated graph to a series of two-qubit operations on the qubits corresponding to the nodes of the graph. The H boxes indicate Hadamard gates, distributed on each qubit. The structure is equivalent to larger multi-qubit operations that select the GHZ state out of the operations. The final state prepared for measurement is shown on the right. For clarity, the displayed graph is visually separable.}
    \label{fig:algorithm-viz}
\end{figure*}

The algorithm consists of a series of two-qubit spiders representing all edges in the graph between the qubits corresponding to the graph nodes. The set of spiders corresponding to a connected graph fragment are equivalent to a fused spider. Since all qubits corresponding to a connected component of the graph are in a GHZ state, upon single qubit measurements they can be measured to be in the same state. 

The contraction of the individual two-qubit spiders can be understood iteratively through the two possible cases. Either two spiders are disconnected and they cannot be contracted. When they do share at least one wire, they can be contracted together to a single spider. It does not matter how many wires connect to the spider, leading to successively larger spiders equivalent to the network of initial two-qubit spiders.

%
%

As the spiders grow, they act on successively larger number of qubits. All the qubits connected to one spider correspond to a connected graph component in the corresponding graph $\mathcal{G}$. Each connected graph fragment thus has one spider connecting all qubits therein as its equivalent contracted statement. 

\subsection{Number of measurements}

We analyse the number of measurements necessary to achieve the correct result within an error bound. 

\subsubsection{Graph connectedness}

After the application of the algorithm, the state is of the form
\begin{align}
    |0_n\rangle \to \bigotimes_k |\text{GHZ}\rangle_k\label{eq:cc-state}
\end{align}
with $k$ components over the $n$ qubits corresponding to connected components of the respective size. The resulting state can be measured. Each state fragment corresponding to a connected component is independent from each other and can be individually measured to be in the ground or excited state. As the probabilities are independent of each other, it is possible to differentiate the different fragments by measurement of all qubits simultaneously. The probability that they are not differentiable, meaning that all fragments are in ground state or excited state, decreases with the number of graph fragments and can be stated as

\begin{subequations}
\begin{align}
    p(\text{disconnected},\text{ failure}) &= \frac{1}{k^M}\\
    p(\text{disconnected}, \text{ success}) &= 1 - \frac{1}{k^M}
\end{align}
\end{subequations}
with $M$ the number of measurements. In the worst case of two graph fragments, the connectedness can be determined after two measurements with probability $p(\text{success}) = \frac34$. 

Correspondingly, if there is only a single graph fragment and the graph is connected, the qubits always measure with the same outcome. Now the probabilities reverse, as the algorithm distinguishes between a connected graph where all qubits are part of the GHZ state and the case where it is disconnected. 

\begin{subequations}
\begin{align}
    p(\text{connected}, \text{ failure}) &= \frac1{2^M}\\
    p(\text{connected}, \text{ success}) &= 1 - \frac{1}{2^M}
\end{align}
\end{subequations}

With two measurements, the probability of success already is at $p(\text{success}) = \frac34$, indicating that the same constant number of measurements is sufficient to distinguish connected and disconnected graphs.

This measurement behaviour allows us to determine the graph connectedness problem in constant time $\mathcal{O}(1)$ of two measurements, independent of the number of graph fragments or number of nodes present in the graph. 

The alternative quantum algorithms \cite{durr_quantum_2006} employs amplitude amplification \cite{brassard_quantum_2002} as a generalization of Grover's algorithm \cite{grover_fast_1996}. Using the speed-up provided by Grover's algorithm, the proposed quantum matrix query algorithm searches the connectivity matrix repeatedly to find edges that are not yet part of the minimum spanning tree. The faster array query relies on an array structure rather than a list to efficiently query the nodes connected to the current node under inspection. The quantum speed-up is achieved by searching through the connected components faster to construct the minimum spanning tree of the graph. These two quantum algorithms are not applicable to finding connected components, as they do not terminate if the graph is not connected. The different implementations, their runtime and required number of operators is listed in table \ref{tab:runtime-connected}.

The two alternative algorithms of \cite{durr_quantum_2006} are hybrid approaches, in which part of a classical algorithm is replaced by a quantum equivalent. Since their algorithms contain searches over the nodes connected to the current node, they achieve the $\mathcal{O}(\sqrt{n})$ speedup to come from the classical $\mathcal{O}(n^3)$ and $\mathcal{O}(n^2)$ runtimes to their achieved $\mathcal{O}(n^{3/2})$ and $\mathcal{O}(n)$ for the matrix and array query model respectively.

\begin{table}
    \centering
    \caption{Overview of the runtime and number of processors for different graph connectedness algorithms.}
\begin{tabular}{lrrr}
Algorithm             & Runtime & \# processors\\\hline
Breadth first search  & $\mathcal{O}(m)$            & 1\\
Parallel connectivity \cite{shiloach_ologn_1982} & $\mathcal{O}(\ln n)$        & $n + 2m$  \\\hline
Matrix query \cite{durr_quantum_2006} & $\mathcal{O}(n^{3/2})$ & 1 \\
Array query \cite{durr_quantum_2006} & $\mathcal{O}(n)$ & 1\\
Quantum connectivity & see text            & $1$ \\\hline
\end{tabular}
    \label{tab:runtime-connected}
\end{table}

\subsubsection{Connected components}

The algorithm can be extended to identify connected components as well, in $\mathcal{O}(|\mathcal{G}_k|)$ steps. Each group of qubits corresponding to a connected component are always in the same measured state. As they are independent of each other, it is possible to identify them based on their coordinated measurement results and therefore the connected components $|\mathcal{G}_k|$ of the graph. With increasing number of measurements, the grouping of qubits becomes clear. 

The difference between the changing measurement outcomes can be measured using the exclusive or operation $\veebar$,
\begin{align}
    a \veebar b = \begin{cases}
        1 & a \ne b\\
        0 & a = b
    \end{cases}
\end{align}
With two successive measurements and the exclusive or operation between them, it is possible to categorize all bits marked with $1$ as belonging to two different GHZ states and therefore different connected components. With repeated measurements, the groups of qubits and therefore the connected components can be found in $M=2|\mathcal{G}_k|$ measurements and therefore in sublinear time. As with the previous approach of detecting graph connectedness, the probability of knowing all connected components correctly is at least $p\ge 3/4$.

\subsection{Depth}\label{ssec:depth}



The algorithm has a variable depth due to the differing number and arrangement of edges in the graph, counted as the number of sequential two-qubit operations to be performed. We assume that two-qubit operations acting on distinct nodes can be parallelized. The best and worst case are given by a brickwork structure of edges and a linear stack, respectively.

\begin{align}
    \text{Best case: }&\mathcal{O} \left(\left\lceil \frac{m}{n} \right\rceil\right)\\
    \text{Worst case: }&\mathcal{O}(m)
\end{align}

for $n$ nodes and $m$ edges of the graph. For a simple graph with no self-loops and no repeating edges, the maximum depth is the number of edges of a complete graph, $m=n(n-1)/2$. The best case correspondingly is $\mathcal{O}(n)$ for a complete graph. The average case, random assignment of two-node edges and asking for the average number of sequential elements, is difficult to calculate analytically. We obtain an analytical upper bound for the average case as

\begin{subequations}
    \begin{align}
    \text{upper bound: } \mathcal{O}(\bar{d}) \\\bar{d} = \frac{\sum_d d \sum_\lambda\sum_i \left(\begin{smallmatrix}
    n \\ 2\lambda_i
\end{smallmatrix}\right)}{\sum_d \sum_\lambda \sum_i \left(\begin{smallmatrix}
    n\\2\lambda_i
\end{smallmatrix}\right)}
    \end{align}
\end{subequations}

based on the integer partitioning without nested edges. $\lambda$ refers to the partitions and sums to $m$, $d$ is a counter over all possible depths and $\bar{d}$ the average depth of the stacked edges. A full derivation is given in appendix \ref{ssec:upper-bound}. We find a numerical lower bound as 

\begin{equation}
    \text{lower bound: }\mathcal{O}\left(\frac{2m}{(1+0.5n)^{0.88}}\right)
\end{equation}

We provide an extensive analysis of the average depth of the circuit in appendix \ref{sec:circuit-depth}. 

Since the two-qubit spiders between qubits corresponding to connected components are contractable to a single spider, one may argue that the depth is trivially $1$, as it is a single multi-qubit operation. However, constructing the large spider explicitly requires the knowledge of which qubits are connected, whereas the two-qubit spider implementation does not require that knowledge.

For sorted graphs such as these commonly generated by programming frameworks, the depth has a upper bound of $\mathcal{O}(2n)$. This low bound can be explained by the fact that the edges can be nested within each other. An example is given in appendix \ref{ssec:sorted-graphs}.






\subsection{State decay}\label{ssec:state-decay}

The projective non-unitary nature of the operations in the proposed quantum algorithm lead to a decay in the quantum state, as information is removed and dissipated to the environment. In the direct naive implementation, this leads to a decay as $1/(2^n-2)$, where in the worst case only two entries in the state vector, expressing the GHZ state, remain. This decay in turn necessitates exponentially many measurements to find the state exactly. This assumes that a dissipated state can be differentiated from a $|0_n\rangle$ state.

The selective nature of the X and Z spider gates can be compared with post-selection techniques used in state preparation \cite{knill_quantum_2002}, quantum cryptography \cite{christandl_postselection_2009, renner_simplifying_2010} and recently in quantum algorithms \cite{wright_automatic_2021}. In these cases the selection removes some of the information of the state. The alternative interpretation is to understand the gate as a stabilizer \cite{poulin_stabilizer_2005, nielsen_quantum_2010}. Stabilizers $\Pi$ are defined as the operations that leave a particular state invariant,
\begin{equation}
    |\psi\rangle = \Pi|\psi\rangle
\end{equation}
The Z spider used here as the two-qubit non-unitary gate acts on the GHZ state in the same way, in the sense that
\begin{equation}
    |\text{GHZ}\rangle = \text{Z spider}(\alpha)|\text{GHZ}\rangle
\end{equation}
up to the phase $\alpha$. Whereas stabilizers due to their unitary nature keep all other states alive, at most transforming them, the non-unitary gate dissipates their information to the open environment and that part of the state is lost. Ignoring this information loss can lead to nonsensical results. It is possible to construct quantum algorithms that solve \textsf{NP} problems with a polynomial runtime, if the state decay is ignored \cite{abrams_nonlinear_1998}.

Obviously this dissipation presents a significant downside to the algorithm. It is however possible to approximate non-unitary transformations with unitary transformations and preserving non-unitary operations such as CNOT gates with the integration of ancilla qubits \cite{terashima_nonunitary_2005}. How many ancilla qubits are necessary to avoid a state decay is left as an open question. We can however determine the worst case by implementing the effects of a Z spider operation using a Bell measurement. The transformation can be understood as the standard way of creating a Bell pair from a GHZ state \cite{nielsen_quantum_2010}.

\begin{figure}[h!]
	$\raisebox{-0.5em}{\hbox{\includegraphics{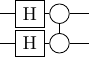}}}
    \Rightarrow
	\raisebox{-2em}{\hbox{\includegraphics{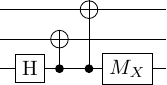}}}$
	\caption{A possible translation from the non-unitary spider to unitary operations together with a measurement. The $M_X$ box indicates a measurementi in the X basis, such that the state on the two continuing qubits is a Bell state, $|00\rangle \pm |11\rangle$.}\label{fig:translation}
\end{figure}

This is shown in figure \ref{fig:translation}. The measurement in the X basis projects the two other qubits of the GHZ state to a Bell state $|00\rangle \pm |11\rangle$, without loss of information to the outside environment. Measuring in the Z basis removes the entanglement \cite{nielsen_quantum_2010}.

An upper limit on the number of necessary ancillas can be given by constructing each two-qubit gate as a three-qubit pair of the two initial qubit together with an ancilla qubit. When the ancilla is measured in the GHZ basis, the initial qubits also flip into the GHZ states. This approach retains fidelity for the two-qubit operation and prevents state decay, at the cost of additional qubits. Hence there is a trade-off of using $\mathcal{O}(2^n)$ calls to the algorithm for the direct two-qubit spider operation with $n$ qubits involved or $\mathcal{O}(1)$ calls but with $m$ ancilla qubits involved, with a worst case total size of the algorithm as $\mathcal{O}(n+m)$.

\subsection{Runtime}

As discussed in the preceding sections, there are several aspects involved in the runtime of the graph connectedness algorithm. The number of measurements is constant, requiring only two measurements to decide whether a graph is connected, $\mathcal{O}(1)$ and a sub-linear number of measurements for identifying connected components, $\mathcal{O}(|\mathcal{G}_k|)$.

The depth of the circuit in both cases depends on the number of edges present in the graph. Some edges can be parallelized, so in terms of parallel steps, only the worst case requires a linear depth of $m$ operations. Overall, the number of parallel and serial operations used the algorithm is always $m$, as the algorithm constructs the graph explicitly.

The algorithm acts as a projector to a specific state, the GHZ state for each connected component of the graph. The remainder of the state dissipates out into the open quantum system and it requires an exponential amount of runs to collect the cases where the quantum state is preserved in the operation of the algorithm. This behaviour can be avoided by adding ancilla qubits. The upper limit of ancillas necessary to preserve the state and prevent dissipation is the number of edges $m$, for total cost of qubits as $\mathcal{O}(n+m)$. The different elements of the runtime are listed in table \ref{tab:runtime-overview}.

\begin{table}[h]
    \centering
    \caption{Overview of the different components of the runtime, with and without ancilla qubits.}
    \begin{tabular}{l|r r}
         & Without ancillas & With ancillas \\\hline
        Measurements &  $\mathcal{O}(2^n) $ & $\mathcal{O}(1)$\\
        Number of qubits & $\mathcal{O}(n)$ & $\mathcal{O}(n + m)$\\
        Depth & $\mathcal{O}(d)$ & $\mathcal{O}(d)$\\\hline
        Time $\times$ space &  $\mathcal{O}(2^n) \times \mathcal{O}(n)$ & $\mathcal{O}(1) \times \mathcal{O}(n+m)$\\\hline
    \end{tabular}
    \label{tab:runtime-overview}
\end{table}

\subsection{Experimental verification}

To best of our knowledge, there is no quantum computer that implements the required non-unitary operation. However, there are a number of experiments that create large GHZ states. Experimentally, a the creation of GHZ state sizes has been pushed in recent years \cite{song_generation_2019, mooney_generation_2021, mooney_whole-device_2021} and experiments with 1000's of qubits have been proposed \cite{zhao_creation_2021}. In the case of applications on commercial quantum computers, the GHZ state is generally created using CNOT and CZ operations. The large scale of the experiments show that the result of our algorithm is generally achievable even on near-term hardware.

Measurement-based quantum computing is non-unitary in its calculations due to its projective measurements \cite{briegel_measurement-based_2009, kissinger_universal_2019}. ZX calculus is used as a language of understanding measurement-based quantum computing and is a natural description of the processes within \cite{duncan_rewriting_2010}. In the literature in the scope of this work, there is no reference to a direct implementation of a X or Z spider as used in our algorithm. However, any non-unitary operation can be achieved using unitary operations and projective measurements \cite{terashima_nonunitary_2005}. The construction of the replacement operation relies on the singular value decomposition, yielding a unitary transformation that needs to be adjusted with $n$-control CNOT gates and ancilla qubits in the $|0\rangle$ state. 

The construction of GHZ states via a Bell measurement on a third qubit has been proposed and experimentally verified on quantum computing hardware \cite{beckey_multipartite_2023}. The construction of the algorithm with ancilla qubits is therefore possible on current hardware already.

Key to the construction of the algorithm is the contractive nature of the spiders from ZX-calculus. A replacement with the similar-looking CZ is not possible, as they lack the key contractive feature.

\section{Discussion}\label{Discussion}

We provide a new quantum algorithm for graph connectedness and connected components. that does not rely on unitary gates, instead working with non-unitary gates to implement the requirements of the problem. The algorithm performs with a constant number of measurements $\mathcal{O}(1)$, needing only two measurements to compute the result with a certainty of $p=3/4$. It exhibits a variable depth depending on the exact graph structure. Optimization can reduce the depth of the circuit, but this optimization on a classical computer likely requires more steps than solving the problem classically. 

The algorithm relies on non-unitary gates. This is necessary to create a large  GHZ state from the ground state with abelian two-qubit operations. Non-unitary operations generally exhibit an probabilistic success pattern. In the present case the possible state space is restricted and the operation can be executed with certainty. The non-unitary gates can implement the graph information and, due to their nilpotent nature, ignore multiple edges.

As the non-unitary gates act as a projector, the state decays compared to its initial state after the application of Hadamard gates on all qubits. This state decay either requires an exponential number of measurements as the number of qubits increases or necessitates the use of ancilla qubits. How many ancillas are necessary is unknown. Without ancilla reuse, the upper bound for ancillas is the number of edges, $m$. 

The extension to non-unitary gates with certain execution on a restricted state space may pave the way to other novel quantum algorithms. Problems on graphs are a natural candidate to this extended quantum approach, as their binary structure is well suited to a reduced state space representation. For graph connectedness and graph components in particular, the two-qubit spiders exhibit the fusing patterns that allows to contract them to an equivalent statement for each connected component.

We believe that the approach of defining the requirements for the algorithm and then considering the necessary requirements for the gates is a helpful tool in constructing new algorithms. In the present case, the fact that the algorithm cannot know the order of the edges leads to the requirement of abelian edges. As we consider repeated edges, the edges need to be nilpotent, which places them outside the framework of unitary operations. Graph problems are particularly suited to this approach, as their discrete nature of nodes maps well to qubits, while the edges can be constructed using two-qubit operations. 

The current structure of the edge gates limits the applicability of the algorithm to contractible global statements about graphs. For other graph problems, the edges would have to be reconsidered to represent the structure necessary for solving the graph problem. The next more difficult graph problems such as bipartiteness or strong connectivity \cite{gross_handbook_2004} have different requirements on the global structure. We believe it is possible to find non-unitary gates that fulfil the global requirements to solve the problem.

\begin{acknowledgments}

MBM acknowledges funding from the German Federal Ministry of Education and Research (BMBF) under the funding program ``Förderprogramm Quantentechnologien – von den Grundlagen zum Markt'' (funding program Quantum Technologies – From Basic Research to Market), projects  BAIQO, 13N16089.

C.F.K. acknowledges financial support by the National Natural Science Foundation of China (Grant nos. 12104524).
\end{acknowledgments}

\appendix

\section{Simulation results}

As the necessary operations are not currently implemented in a quantum computer, we show the structure of the algorithm in simulation. Any state simulation-based program should be able to perform the necessary calculation. We implement our system in Julia \cite{bezanson2017julia} and its quantum simulation package Yao \cite{luo_yaojl_2020}. In the supplementary material, we provide a notebook that implements an interface between a graph and the quantum algorithm. The circuit is implemented without ancilla qubits and ignores state decay, as measurement samples are taken from the remaining state. The decay can be observed when inspecting the state.

With the supplementary material, it is possible to verify the implementation of the algorithm. We structure it such that particular graphs can be loaded and converted into quantum circuits. The alternative is to use the random Erdős-Rényi graphs \cite{erdos_random_1959} already set up in the notebook.

Measurements of the state after the algorithm and the verification of the output against the true connectedness is independent. New graphs and new measurements can be generated by repeatedly executing the relevant cell of the notebook. The number of shots is an optional parameter.

For the graph presented in figure \ref{fig:algorithm-viz}, there are four different measurement results with equal probability. These are
\begin{equation}
    \text{outcome} = \begin{cases}
        000000000\\
        000111111\\
        111000000\\
        111111111\\
    \end{cases}
\end{equation}

With the minimum amount of two measurements, two of the outcomes are selected at random. It is apparent that the combined probability of detecting the disconnected nature of the graph is $p_{M=2} = 3/4$.

\section{Quantum Circuit Depth}\label{sec:circuit-depth}

In principle, the number of edges in series can be calculated by summing over combinatorial permutations of edges. This is possible in numerical simulation, but the analyical approach quickly fails as the number of options explodes. We can analytically derive an upper bound by looking at the problem as an integer partition problem \cite{mertens_easiest_2005}. In this case the nesting of the edges within each other is ignored, leading to a strict upper bound to the depth.



\subsection{Upper bound as an integer partition problem}\label{ssec:upper-bound}

To derive an analytical upper bound on the depth of the quantum circuit, the original problem can be reformulated as a combinatorial problem. We can introduce a matrix with $d$ rows and $n^\prime\equiv n-1$ columns, where each element is either $0$ or $1$. The objective is to compute the connected components of $1$s within each row. For instance, for $n=10$ and $d=9$, a typical matrix takes the following form
\begin{equation}
    \begin{pmatrix}
        1 & 0 & 1 & 0 & 1 & 1 & 1 & 1 & 1\\
        0 & 1 & 0 & 1 & 0 & 1 & 1 & 0 & 0 \\
        1 & 1 & 1 & 1 & 0 & 0 & 1 & 1 & 1\\
        1 & 1 & 1 & 1 & 1 & 0 & 1 & 1 & 0 \\
        0 & 1 & 1 & 1 & 1 & 1 & 0 & 0 & 0 \\
        0 & 0 & 1 & 1 & 1 & 1 & 1 & 0 & 0 \\
        0 & 0 & 0 & 1 & 1 & 1 & 1 & 1 & 0 \\
        0 & 0 & 0 & 0 & 0 & 1 & 1 & 1 & 0 \\
        0 & 0 & 0 & 0 & 0 & 0 & 0 & 1 & 1
    \end{pmatrix}.
\end{equation}
For example, the connected components of 1s within the first row are two $[1]$ and one $[1,1,1,1,1]$. Thus, the first row has three connected components of 1s. In general, one can denote the number of connected components of 1s in the $i$-th row as $\lambda_i$, which are subjected to the constraint
\begin{equation}
    \lambda_1+\lambda_2+\cdots+\lambda_d = m.
\end{equation}
From this one can see that the tuple $\lambda\equiv (\lambda_1,\lambda_2,\cdots,\lambda_d)$ forms a partition of $m$ into exactly $d$ parts, where $m$ is the total number of connected components of 1s in the matrix. Here, the number $m$ is also the number of edges in the original graph problem. In the above example, we have $\lambda=(3,3,2,2,1,1,1,1,1)$, which is a partition of $m=15$ into exactly $d=9$ parts. Now, the problem becomes: what is the number $N(n^\prime,\lambda)$ of possible ways to fill the 0s and 1s in a row of length $n^\prime$, such that there is exactly $\lambda$ connected components of 1s in that row? Some observations lead to the following recurrence relation
\begin{align}
    N(n^\prime,\lambda)&=2N(n^\prime-1,\lambda)-N(n^\prime-2,\lambda)\nonumber\\
    &+N(n^\prime-2,\lambda-1).
\end{align}
A table for $N(n^\prime,\lambda)$ for $n^\prime\leq 13$ is given in table \ref{tab:nnlambda}.

\begin{table}[h]
\caption{The structure of possible partitions $\lambda$ within a row of length $n'$. The last line indicates the sum over the row.}
\begin{tabular}{c|ccccccccccccc}
\hline
$\lambda \backslash n^\prime$ & 1 & 2 & 3 & 4 & 5 & 6 & 7 & 8 & 9 & 10 & 11 & 12 & 13 \\
\hline
0 & 1 & 1 & 1 & 1 & 1 & 1 & 1 & 1 & 1 & 1 & 1 & 1 & 1 \\
1 & 1 & 3 & 6 & 10 & 15 & 21 & 28 & 36 & 45 & 55 & 66 & 78 & 91 \\
2 & 0 & 0 & 1 & 5 & 15 & 35 & 70 & 126 & 210 & 330 & 495 & 715 & 1001 \\
3 & 0 & 0 & 0 & 0 & 1 & 7 & 28 & 84 & 210 & 462 & 924 & 1716 & 3003 \\
4 & 0 & 0 & 0 & 0 & 0 & 0 & 1 & 9 & 45 & 165 & 495 & 1287 & 3003 \\
5 & 0 & 0 & 0 & 0 & 0 & 0 & 0 & 0 & 1 & 11 & 66 & 286 & 1001 \\
6 & 0 & 0 & 0 & 0 & 0 & 0 & 0 & 0 & 0 & 0 & 1 & 13 & 91 \\
7 & 0 & 0 & 0 & 0 & 0 & 0 & 0 & 0 & 0 & 0 & 0 & 0 & 1 \\
\hline
$\sum$ & $2^1$ & $2^2$ & $2^3$ & $2^4$ & $2^5$ & $2^6$ & $2^7$ & $2^8$ & $2^9$ & $2^{10}$ & $2^{11}$ & $2^{12}$ & $2^{13}$ \\
\hline
\end{tabular}
\label{tab:nnlambda}
\end{table}

The bottom of the table indicates the sum of all $N(n^\prime,\lambda)$ for each $n^\prime$. Clearly, the number $N(n^\prime,\lambda)$ satisfy the rule:

\begin{equation}
\sum_{\lambda=0}^{\left\lceil \frac{n^\prime}{2} \right\rceil} N(n^\prime,\lambda) = 2^{n^\prime},
\end{equation}

where $\left\lceil \frac{n^\prime}{2} \right\rceil$ is the ceiling function for $\frac{n^\prime}{2}$. At first glance, this recurrence relation appears hard to solve. But it becomes easier upon noticing that for $\lambda=1$, one can explicitly write the formula $N(n^\prime,1)=\binom{n^\prime+1}{2}$. Thus, the recurrence relation for $\lambda=2$ can be rewrite as

\begin{equation}
    c_{n^\prime}-c_{n^\prime-1}-(c_{n^\prime-1}-c_{n^\prime-2})=\binom{n^\prime-1}{2},
\end{equation}

where $c_{n^\prime}\equiv N(n^\prime,2)$. A direct computation yields

\begin{equation}\label{recurrence}
    c_{n^\prime}-c_{n^\prime-1}=\sum_{k=1}^{n^\prime-2}\binom{n^\prime-k}{2}=\binom{n^\prime}{3},
\end{equation}

where we have used the initial condition that $c_1=c_2=0$. Interestingly, the sum of binomial coefficients might have a combinatorial Interpretation: the sum of these binomial coefficients is essentially counting the number of ways to select $2$ elements from sets of size $n^\prime-1$, $n^\prime-2$, $\cdots$, $1$. This is equivalent to counting the number of ways to select $3$ elements from sets of size $n^\prime$. From equation \eqref{recurrence}, one can immediately deduce the formula

\begin{equation}
    N(n^\prime,2)\equiv c_{n^\prime}=\sum_{k=0}^{n^\prime-3}\binom{n^\prime-k}{3}=\binom{n^\prime+1}{4}.
\end{equation}

By applying similar methods, one can derive the general formula for arbitrary $\lambda$:

\begin{equation}
    N(n^\prime,\lambda)= \binom{n^\prime+1}{2\lambda}.
\end{equation}

Hence, for a given Young diagram corresponding to a restricted partition of $m$, the total number of connected components of 1's in all $d$ rows with length $n^\prime=n-1$ is

\begin{equation}
    N_\lambda = \sum_{i=1}^d\binom{n}{2\lambda_i},
\end{equation}

where $\lambda_i$ are subjected to the constraint $\lambda_1+\lambda_2+\cdots+\lambda_d=m$. Hence, for a given depth $d$, the total number $N_d$ is the sum of all $N_\lambda$ for different Young diagrams, where the order of the summands is considered distinct. Explicitly, the total number $N_d$ is given by

\begin{equation}
    N_d=\sum_{\lambda}N_\lambda = \sum_{\lambda}\sum_i\binom{n}{2\lambda_i}.
\end{equation}

For example, the three partitions $2+1+1$, $1+2+1$ and $1+1+2$, all correspond to the same Young diagram, but they are considered distinct because they are different compositions of the number $4$ into $3$ parts. Finally, the mean value of the depth $d$ is computed by

\begin{equation}
    \bar{d} \equiv \frac{\sum_d d N_d}{\sum_d N_d}=\frac{\sum_dd\sum_\lambda\sum_i\binom{n}{2\lambda_i}}{\sum_d\sum_\lambda\sum_i\binom{n}{2\lambda_i}}.
\end{equation}

This formula provides an analytical upper bound for the average depth $d$ of the quantum circuit. 

\begin{figure}[h]
%
%
%
%
%
%
\includegraphics{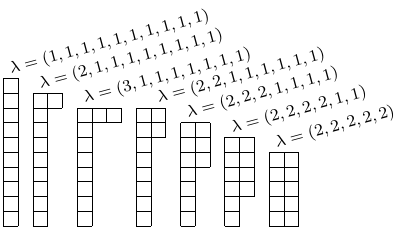}
\caption{The Young diagrams used to calculate the example of $n=5$ and $m=10$.}\label{fig:young-diagrams}
\end{figure}

For example, when $n=5$ and $m=10$, the largest possible depth is $d=10$. In this case, the only possible partition of $m=10$ is $\lambda=(1, 1, 1, 1, 1, 1, 1, 1, 1, 1)$. Then one obtains $N_{10}=10\times \binom{5}{2}=100$. For $d=9$, the only possible partition of $m=10$ is $\lambda=(2, 1, 1, 1, 1, 1, 1, 1, 1)$ Then one obtains $N_9=9\times(\binom{5}{4}+8\times\binom{5}{2})=765$. 

For $d=8$, the two possible partition of $m=10$ is $\lambda=(3,1,1,1,1,1,1,1)$ and $(2,2,1,1,1,1,1,1)$. Due to the restriction $2\lambda_i\leq n$ for all $\lambda_i$, the largest possible value of $\lambda_i$ should be $\left\lfloor \frac{n}{2} \right\rfloor$. In our case, the largest $\lambda_i$ is $2$. Hence, the first Young diagram does not contribute. Counting the second Young diagram, we obtain $N_8=\binom{8}{2}\times (2\times \binom{5}{4}+6\times \binom{5}{2})=1960$.

Using the same logic, for $d=7$, although there are 3 partitions, the only relevant partition for $m=10$ is $\lambda=(2,2,2,1,1,1,1)$. Then one obtains $N_7=\binom{7}{3}\times(3\times\binom{5}{4}+4\times\binom{5}{2})=1925$. For $d=6$, while there are a total 5 partitions,  the only  relevant partition for $m=10$ is $\lambda=(2,2,2,2,1,1)$.

Then one obtains $N_6=\binom{6}{4}\times (4\times \binom{5}{4}+2\times \binom{5}{2})=600$. Finally, for $d=5$, while there are a total of 7 partitions, the only relevant partition for  $m=10$ is $\lambda=(2,2,2,2,2)$. All Young diagrams mentioned here are drawn in figure \ref{fig:young-diagrams}.

\begin{figure}[ht]
\centering
	\includegraphics{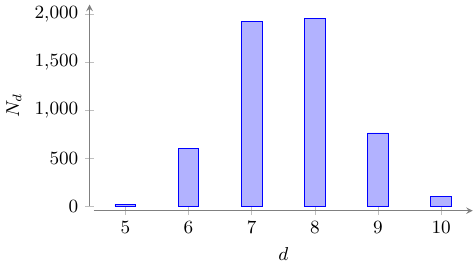}
\caption{Histogram of the distribution of depths in the quantum circuit based on the integer partition for $n=5$ and $m=10$.}\label{fig:histogram} 
\end{figure}

As such, one obtains $N_5=5\times\binom{5}{4}=25$. The statistics for the upper bound of the quantum circuit are shown in Fig.\:\ref{fig:histogram}.

From this, the weighted mean of $d$ is immediately calculated as $\bar{d}=8.14$, which gives the upper bound for the depth of an arbitrary graph with $n=5$ vertices and $m=10$ edges. This example clearly demonstrates that the number of Young diagrams used to calculate the upper bound of the quantum circuit depth grows rapidly, unless $n$ is significantly smaller than $m$. The upper bound on the number of Young diagrams required for each d is given by the restricted partition number of m. For example, the restricted partition number of 10 into 5 parts is 7.

In general, the number of partitions of $n$ into exactly $k$ parts, denoted as $p_k(n)$, is given by table \ref{tab:k-partitions}, which summarizes the cases for $n\leq 12$.

\begin{table}
\caption{The number of partitions of $n$ into $k$ parts, up to $n = 12$. The number of partitions does not follow a clear structure.}\label{tab:k-partitions}
\begin{tabular}{c|*{12}{p{1em}}}
\hline
$n \backslash k$ & 1 & 2 & 3 & 4 & 5 & 6 & 7 & 8 & 9 & 10 & 11 & 12 \\
\hline
1  & 1 & 0 & 0 & 0 & 0 & 0 & 0 & 0 & 0 & 0 & 0 & 0 \\
2  & 1 & 1 & 0 & 0 & 0 & 0 & 0 & 0 & 0 & 0 & 0 & 0 \\
3  & 1 & 1 & 1 & 0 & 0 & 0 & 0 & 0 & 0 & 0 & 0 & 0 \\
4  & 1 & 2 & 1 & 1 & 0 & 0 & 0 & 0 & 0 & 0 & 0 & 0 \\
5  & 1 & 2 & 2 & 1 & 1 & 0 & 0 & 0 & 0 & 0 & 0 & 0 \\
6  & 1 & 3 & 3 & 2 & 1 & 1 & 0 & 0 & 0 & 0 & 0 & 0 \\
7  & 1 & 3 & 4 & 3 & 2 & 1 & 1 & 0 & 0 & 0 & 0 & 0 \\
8  & 1 & 4 & 5 & 5 & 3 & 2 & 1 & 1 & 0 & 0 & 0 & 0 \\
9  & 1 & 4 & 7 & 6 & 5 & 3 & 2 & 1 & 1 & 0 & 0 & 0 \\
10 & 1 & 5 & 8 & 9 & 7 & 5 & 3 & 2 & 1 & 1 & 0 & 0 \\
11 & 1 & 5 & 10 & 11 & 10 & 7 & 5 & 3 & 2 & 1 & 1 & 0 \\
12 & 1 & 6 & 12 & 15 & 13 & 11 & 7 & 5 & 3 & 2 & 1 & 1 \\
\hline
\end{tabular}
\end{table}

\subsection{Numerical approximation}\label{ssec:numerical-solution}

\begin{figure}
%
	\includegraphics{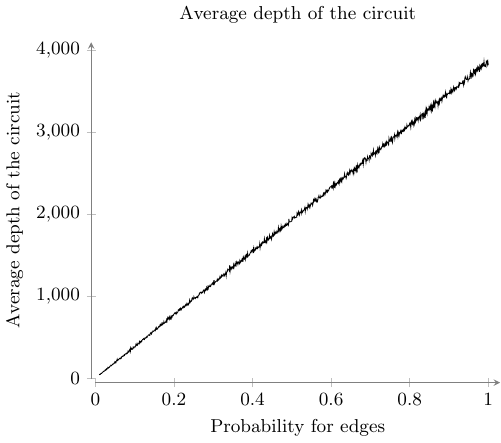}
\label{fig:growth-of-m}\caption{The depth of the circuit as the probability of an edge in an Erdōs-Rényi graph increases, for $n=1000$. Edge assignment is randomized. The data shows the average depth over 10000 iterations, in 1000 steps of edge probability from $p_\text{min} = 0.01$ to $p_\text{max} = 1$. The number of edges in the graph scales linearly with the probability. Noise on the line stems from the depth variations.}
\end{figure}

A numerical approach is straightforward. For simple graphs, the standard graph analysis packages can be used to generate random Erdős-Rényi graphs \cite{erdos_random_1959}, which can then be assigned to the nodes. Erdős-Rényi graphs have a probability of connectivity, where is each edge is either connected or not based on this probability. Many graph packages output a sorted list of edges, leading to the low worst case bound illustrated in the following section \ref{ssec:sorted-graphs}. If the list of edges is shuffled, the desired random assignment behavior occurs. An example for $n=1000$ is shown in figure \ref{fig:growth-of-m}. The probability of an edge linearly correlates with the number of edges in the graph, and this is linear in turn with the depth of the circuit. For large numbers $n$, it also shows that the random approach is already efficient, the 499,500 edges present in the complete in the graph correspond to a depth of the quantum circuit of less than 4000 sequential two-qubit operations on average.

\begin{figure}
%
	\includegraphics{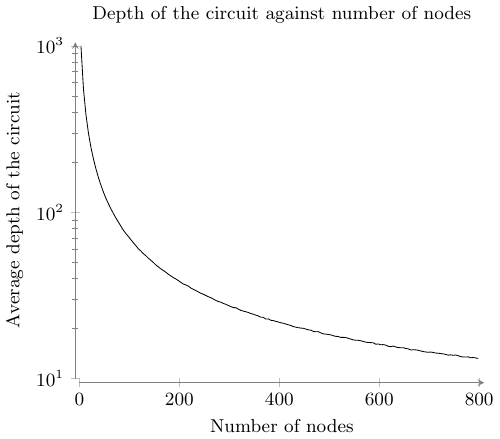}
\caption{As the number of nodes increases, the average depth decreases for a constant number of $m$. This graph shows the average depth of the circuit for $m=1000$, averaged over 500 iterations for a node count from $n_\text{min}=3$ to $n_\text{max}=800$ in steps of 5. The $y$ axis is logarithmic, which illustrates that the behaviour does not follow a simple exponential relationship.}\label{fig:depth-vs-n}
\end{figure}

The other part of the depth is the number of nodes. A plot illustrating the change in depth with increasing number of nodes is shown in figure \ref{fig:depth-vs-n}. As the number of nodes increases for a constant $m$, the depth decreases. It is not a simple exponential relationship, as can be read from the figure. The best experimental fit is one of the form 

\begin{equation}
    \text{lower bound: } \mathcal{O}\left(\frac{2m}{(1+.5 n)^{.88}}\right)
\end{equation}. 

This fit still underestimates the behavior for large $n$ but provides a reasonable lower bound.

\subsection{Sorted graphs}\label{ssec:sorted-graphs}

\begin{figure}
    \centering
%
%
%
	\includegraphics{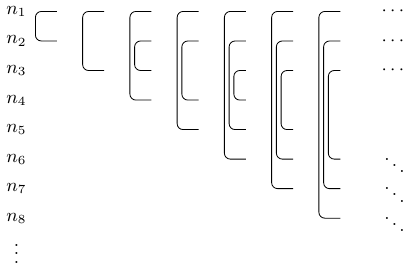}
    \caption{Visual explanation of the graph stacking as discussed in section \ref{ssec:sorted-graphs}. For a sorted arrangement of edges, the later edges can be nested between the previous nodes. Different dot and dash pattern correspond to the different initial node of the edge.}
    \label{fig:graph-stacking}

\end{figure}

For sorted graphs, the upper bound in depth with no repeated edges and no self-loops is $\mathcal{O}(2n)$, due to the edges stacking within each other. This is explained visually in figure \ref{fig:graph-stacking} for a fraction of a complete graph. Each subsequent series of edges nests with the previous layer, increasing the overall stack height only by 2. As explained in the main the text, sorting the list of edges to achieve this low bound using a classical sorting algorithm requires $\mathcal{O}(m \log m)$ operations, longer than the standard breadth first search approach of classically solving the problem.

\input{main.bbl}
\end{document}

%% file: main.bbl
%

%% file: main.bbl
\begin{thebibliography}{60}%
\makeatletter
\providecommand \@ifxundefined [1]{%
 \@ifx{#1\undefined}
}%
\providecommand \@ifnum [1]{%
 \ifnum #1\expandafter \@firstoftwo
 \else \expandafter \@secondoftwo
 \fi
}%
\providecommand \@ifx [1]{%
 \ifx #1\expandafter \@firstoftwo
 \else \expandafter \@secondoftwo
 \fi
}%
\providecommand \natexlab [1]{#1}%
\providecommand \enquote  [1]{``#1''}%
\providecommand \bibnamefont  [1]{#1}%
\providecommand \bibfnamefont [1]{#1}%
\providecommand \citenamefont [1]{#1}%
\providecommand \href@noop [0]{\@secondoftwo}%
\providecommand \href [0]{\begingroup \@sanitize@url \@href}%
\providecommand \@href[1]{\@@startlink{#1}\@@href}%
\providecommand \@@href[1]{\endgroup#1\@@endlink}%
\providecommand \@sanitize@url [0]{\catcode `\\12\catcode `\$12\catcode
  `\&12\catcode `\#12\catcode `\^12\catcode `\_12\catcode `\%12\relax}%
\providecommand \@@startlink[1]{}%
\providecommand \@@endlink[0]{}%
\providecommand \url  [0]{\begingroup\@sanitize@url \@url }%
\providecommand \@url [1]{\endgroup\@href {#1}{\urlprefix }}%
\providecommand \urlprefix  [0]{URL }%
\providecommand \Eprint [0]{\href }%
\providecommand \doibase [0]{https://doi.org/}%
\providecommand \selectlanguage [0]{\@gobble}%
\providecommand \bibinfo  [0]{\@secondoftwo}%
\providecommand \bibfield  [0]{\@secondoftwo}%
\providecommand \translation [1]{[#1]}%
\providecommand \BibitemOpen [0]{}%
\providecommand \bibitemStop [0]{}%
\providecommand \bibitemNoStop [0]{.\EOS\space}%
\providecommand \EOS [0]{\spacefactor3000\relax}%
\providecommand \BibitemShut  [1]{\csname bibitem#1\endcsname}%
\let\auto@bib@innerbib\@empty
\bibitem [{\citenamefont {Nielsen}\ and\ \citenamefont
  {Chuang}(2010)}]{nielsen_quantum_2010}%
  \BibitemOpen
  \bibfield  {author} {\bibinfo {author} {\bibfnamefont {M.~A.}\ \bibnamefont
  {Nielsen}}\ and\ \bibinfo {author} {\bibfnamefont {I.~L.}\ \bibnamefont
  {Chuang}},\ }\href@noop {} {\emph {\bibinfo {title} {Quantum computation and
  quantum information}}},\ \bibinfo {edition} {10th}\ ed.\ (\bibinfo
  {publisher} {Cambridge University Press},\ \bibinfo {address} {Cambridge ;
  New York},\ \bibinfo {year} {2010})\BibitemShut {NoStop}%
\bibitem [{\citenamefont {Coecke}\ \emph {et~al.}(2016)\citenamefont {Coecke},
  \citenamefont {Fritz},\ and\ \citenamefont
  {Spekkens}}]{coecke2016mathematical}%
  \BibitemOpen
  \bibfield  {author} {\bibinfo {author} {\bibfnamefont {B.}~\bibnamefont
  {Coecke}}, \bibinfo {author} {\bibfnamefont {T.}~\bibnamefont {Fritz}},\ and\
  \bibinfo {author} {\bibfnamefont {R.~W.}\ \bibnamefont {Spekkens}},\
  }\bibfield  {title} {\bibinfo {title} {A mathematical theory of resources},\
  }\href@noop {} {\bibfield  {journal} {\bibinfo  {journal} {Information and
  Computation}\ }\textbf {\bibinfo {volume} {250}},\ \bibinfo {pages} {59}
  (\bibinfo {year} {2016})}\BibitemShut {NoStop}%
\bibitem [{\citenamefont {Schmid}\ \emph {et~al.}(2020)\citenamefont {Schmid},
  \citenamefont {Rosset},\ and\ \citenamefont {Buscemi}}]{schmid2020type}%
  \BibitemOpen
  \bibfield  {author} {\bibinfo {author} {\bibfnamefont {D.}~\bibnamefont
  {Schmid}}, \bibinfo {author} {\bibfnamefont {D.}~\bibnamefont {Rosset}},\
  and\ \bibinfo {author} {\bibfnamefont {F.}~\bibnamefont {Buscemi}},\
  }\bibfield  {title} {\bibinfo {title} {The type-independent resource theory
  of local operations and shared randomness},\ }\href@noop {} {\bibfield
  {journal} {\bibinfo  {journal} {Quantum}\ }\textbf {\bibinfo {volume} {4}},\
  \bibinfo {pages} {262} (\bibinfo {year} {2020})}\BibitemShut {NoStop}%
\bibitem [{\citenamefont {Tilma}\ \emph {et~al.}(2010)\citenamefont {Tilma},
  \citenamefont {Hamaji}, \citenamefont {Munro},\ and\ \citenamefont
  {Nemoto}}]{tilma2010entanglement}%
  \BibitemOpen
  \bibfield  {author} {\bibinfo {author} {\bibfnamefont {T.}~\bibnamefont
  {Tilma}}, \bibinfo {author} {\bibfnamefont {S.}~\bibnamefont {Hamaji}},
  \bibinfo {author} {\bibfnamefont {W.}~\bibnamefont {Munro}},\ and\ \bibinfo
  {author} {\bibfnamefont {K.}~\bibnamefont {Nemoto}},\ }\bibfield  {title}
  {\bibinfo {title} {Entanglement is not a critical resource for quantum
  metrology},\ }\href@noop {} {\bibfield  {journal} {\bibinfo  {journal}
  {Physical Review A—Atomic, Molecular, and Optical Physics}\ }\textbf
  {\bibinfo {volume} {81}},\ \bibinfo {pages} {022108} (\bibinfo {year}
  {2010})}\BibitemShut {NoStop}%
\bibitem [{\citenamefont {Catani}(2018)}]{catani2018non}%
  \BibitemOpen
  \bibfield  {author} {\bibinfo {author} {\bibfnamefont {L.}~\bibnamefont
  {Catani}},\ }\emph {\bibinfo {title} {Non-classicality as a computational
  resource}},\ \href@noop {} {Ph.D. thesis},\ \bibinfo  {school} {UCL
  (University College London)} (\bibinfo {year} {2018})\BibitemShut {NoStop}%
\bibitem [{\citenamefont {Kam}\ \emph {et~al.}(2023)\citenamefont {Kam},
  \citenamefont {Zhang},\ and\ \citenamefont {Feng}}]{kam2023coherent}%
  \BibitemOpen
  \bibfield  {author} {\bibinfo {author} {\bibfnamefont {C.~F.}\ \bibnamefont
  {Kam}}, \bibinfo {author} {\bibfnamefont {W.~M.}\ \bibnamefont {Zhang}},\
  and\ \bibinfo {author} {\bibfnamefont {D.~H.}\ \bibnamefont {Feng}},\
  }\href@noop {} {\emph {\bibinfo {title} {Coherent States: New Insights into
  Quantum Mechanics with Applications}}}\ (\bibinfo  {publisher} {Springer},\
  \bibinfo {year} {2023})\BibitemShut {NoStop}%
\bibitem [{\citenamefont {Horodecki}\ \emph {et~al.}(2009)\citenamefont
  {Horodecki}, \citenamefont {Horodecki}, \citenamefont {Horodecki},\ and\
  \citenamefont {Horodecki}}]{horodecki2009quantum}%
  \BibitemOpen
  \bibfield  {author} {\bibinfo {author} {\bibfnamefont {R.}~\bibnamefont
  {Horodecki}}, \bibinfo {author} {\bibfnamefont {P.}~\bibnamefont
  {Horodecki}}, \bibinfo {author} {\bibfnamefont {M.}~\bibnamefont
  {Horodecki}},\ and\ \bibinfo {author} {\bibfnamefont {K.}~\bibnamefont
  {Horodecki}},\ }\bibfield  {title} {\bibinfo {title} {Quantum entanglement},\
  }\href@noop {} {\bibfield  {journal} {\bibinfo  {journal} {Reviews of modern
  physics}\ }\textbf {\bibinfo {volume} {81}},\ \bibinfo {pages} {865}
  (\bibinfo {year} {2009})}\BibitemShut {NoStop}%
\bibitem [{\citenamefont {Bengtsson}\ and\ \citenamefont
  {{\.Z}yczkowski}(2017)}]{bengtsson2017geometry}%
  \BibitemOpen
  \bibfield  {author} {\bibinfo {author} {\bibfnamefont {I.}~\bibnamefont
  {Bengtsson}}\ and\ \bibinfo {author} {\bibfnamefont {K.}~\bibnamefont
  {{\.Z}yczkowski}},\ }\href@noop {} {\emph {\bibinfo {title} {Geometry of
  quantum states: an introduction to quantum entanglement}}}\ (\bibinfo
  {publisher} {Cambridge university press},\ \bibinfo {year}
  {2017})\BibitemShut {NoStop}%
\bibitem [{\citenamefont {Schmid}\ \emph {et~al.}(2023)\citenamefont {Schmid},
  \citenamefont {Fraser}, \citenamefont {Kunjwal}, \citenamefont {Sainz},
  \citenamefont {Wolfe},\ and\ \citenamefont
  {Spekkens}}]{schmid2023understanding}%
  \BibitemOpen
  \bibfield  {author} {\bibinfo {author} {\bibfnamefont {D.}~\bibnamefont
  {Schmid}}, \bibinfo {author} {\bibfnamefont {T.~C.}\ \bibnamefont {Fraser}},
  \bibinfo {author} {\bibfnamefont {R.}~\bibnamefont {Kunjwal}}, \bibinfo
  {author} {\bibfnamefont {A.~B.}\ \bibnamefont {Sainz}}, \bibinfo {author}
  {\bibfnamefont {E.}~\bibnamefont {Wolfe}},\ and\ \bibinfo {author}
  {\bibfnamefont {R.~W.}\ \bibnamefont {Spekkens}},\ }\bibfield  {title}
  {\bibinfo {title} {Understanding the interplay of entanglement and
  nonlocality: motivating and developing a new branch of entanglement theory},\
  }\href@noop {} {\bibfield  {journal} {\bibinfo  {journal} {Quantum}\ }\textbf
  {\bibinfo {volume} {7}},\ \bibinfo {pages} {1194} (\bibinfo {year}
  {2023})}\BibitemShut {NoStop}%
\bibitem [{\citenamefont {Scarani}(2019)}]{scarani2019bell}%
  \BibitemOpen
  \bibfield  {author} {\bibinfo {author} {\bibfnamefont {V.}~\bibnamefont
  {Scarani}},\ }\href@noop {} {\emph {\bibinfo {title} {Bell nonlocality}}}\
  (\bibinfo  {publisher} {Oxford University Press},\ \bibinfo {year}
  {2019})\BibitemShut {NoStop}%
\bibitem [{\citenamefont {Wolfe}\ \emph {et~al.}(2020)\citenamefont {Wolfe},
  \citenamefont {Schmid}, \citenamefont {Sainz}, \citenamefont {Kunjwal},\ and\
  \citenamefont {Spekkens}}]{wolfe2020quantifying}%
  \BibitemOpen
  \bibfield  {author} {\bibinfo {author} {\bibfnamefont {E.}~\bibnamefont
  {Wolfe}}, \bibinfo {author} {\bibfnamefont {D.}~\bibnamefont {Schmid}},
  \bibinfo {author} {\bibfnamefont {A.~B.}\ \bibnamefont {Sainz}}, \bibinfo
  {author} {\bibfnamefont {R.}~\bibnamefont {Kunjwal}},\ and\ \bibinfo {author}
  {\bibfnamefont {R.~W.}\ \bibnamefont {Spekkens}},\ }\bibfield  {title}
  {\bibinfo {title} {Quantifying bell: The resource theory of nonclassicality
  of common-cause boxes},\ }\href@noop {} {\bibfield  {journal} {\bibinfo
  {journal} {Quantum}\ }\textbf {\bibinfo {volume} {4}},\ \bibinfo {pages}
  {280} (\bibinfo {year} {2020})}\BibitemShut {NoStop}%
\bibitem [{\citenamefont {Rosset}\ \emph {et~al.}(2020)\citenamefont {Rosset},
  \citenamefont {Schmid},\ and\ \citenamefont {Buscemi}}]{rosset2020type}%
  \BibitemOpen
  \bibfield  {author} {\bibinfo {author} {\bibfnamefont {D.}~\bibnamefont
  {Rosset}}, \bibinfo {author} {\bibfnamefont {D.}~\bibnamefont {Schmid}},\
  and\ \bibinfo {author} {\bibfnamefont {F.}~\bibnamefont {Buscemi}},\
  }\bibfield  {title} {\bibinfo {title} {Type-independent characterization of
  spacelike separated resources},\ }\href@noop {} {\bibfield  {journal}
  {\bibinfo  {journal} {Physical Review Letters}\ }\textbf {\bibinfo {volume}
  {125}},\ \bibinfo {pages} {210402} (\bibinfo {year} {2020})}\BibitemShut
  {NoStop}%
\bibitem [{\citenamefont {Montanaro}(2016)}]{montanaro_quantum_2016}%
  \BibitemOpen
  \bibfield  {author} {\bibinfo {author} {\bibfnamefont {A.}~\bibnamefont
  {Montanaro}},\ }\bibfield  {title} {\bibinfo {title} {Quantum algorithms: an
  overview},\ }\href {https://doi.org/10.1038/npjqi.2015.23} {\bibfield
  {journal} {\bibinfo  {journal} {npj Quantum Information}\ }\textbf {\bibinfo
  {volume} {2}},\ \bibinfo {pages} {1} (\bibinfo {year} {2016})},\ \bibinfo
  {note} {number: 1 Publisher: Nature Publishing Group}\BibitemShut {NoStop}%
\bibitem [{\citenamefont {Cormen}\ \emph {et~al.}(2009)\citenamefont {Cormen},
  \citenamefont {Leiserson}, \citenamefont {Rivest},\ and\ \citenamefont
  {Stein}}]{cormen_introduction_2009}%
  \BibitemOpen
  \bibfield  {author} {\bibinfo {author} {\bibfnamefont {T.~H.}\ \bibnamefont
  {Cormen}}, \bibinfo {author} {\bibfnamefont {C.~E.}\ \bibnamefont
  {Leiserson}}, \bibinfo {author} {\bibfnamefont {R.~L.}\ \bibnamefont
  {Rivest}},\ and\ \bibinfo {author} {\bibfnamefont {C.}~\bibnamefont
  {Stein}},\ }\href@noop {} {\emph {\bibinfo {title} {Introduction to
  algorithms}}},\ \bibinfo {edition} {third edition}\ ed.\ (\bibinfo
  {publisher} {MIT Press},\ \bibinfo {address} {Cambridge, Mass.},\ \bibinfo
  {year} {2009})\ \bibinfo {note} {oCLC: 311310321}\BibitemShut {NoStop}%
\bibitem [{\citenamefont {Sipser}(1996)}]{sipser_introduction_1996}%
  \BibitemOpen
  \bibfield  {author} {\bibinfo {author} {\bibfnamefont {M.}~\bibnamefont
  {Sipser}},\ }\href {https://dl.acm.org/doi/10.1145/230514.571645} {\emph
  {\bibinfo {title} {Introduction to the {Theory} of {Computation}}}},\
  Vol.~\bibinfo {volume} {27}\ (\bibinfo {year} {1996})\BibitemShut {NoStop}%
\bibitem [{\citenamefont {Bernstein}\ and\ \citenamefont
  {Vazirani}(1997)}]{bernstein_quantum_1997}%
  \BibitemOpen
  \bibfield  {author} {\bibinfo {author} {\bibfnamefont {E.}~\bibnamefont
  {Bernstein}}\ and\ \bibinfo {author} {\bibfnamefont {U.}~\bibnamefont
  {Vazirani}},\ }\bibfield  {title} {\bibinfo {title} {Quantum {Complexity}
  {Theory}},\ }\href {https://doi.org/10.1137/S0097539796300921} {\bibfield
  {journal} {\bibinfo  {journal} {SIAM Journal on Computing}\ }\textbf
  {\bibinfo {volume} {26}},\ \bibinfo {pages} {1411} (\bibinfo {year}
  {1997})},\ \bibinfo {note} {publisher: Society for Industrial and Applied
  Mathematics}\BibitemShut {NoStop}%
\bibitem [{\citenamefont {Ediger}\ \emph {et~al.}(2010)\citenamefont {Ediger},
  \citenamefont {Jiang}, \citenamefont {Riedy}, \citenamefont {Bader},
  \citenamefont {Corley}, \citenamefont {Farber},\ and\ \citenamefont
  {Reynolds}}]{ediger_massive_2010}%
  \BibitemOpen
  \bibfield  {author} {\bibinfo {author} {\bibfnamefont {D.}~\bibnamefont
  {Ediger}}, \bibinfo {author} {\bibfnamefont {K.}~\bibnamefont {Jiang}},
  \bibinfo {author} {\bibfnamefont {J.}~\bibnamefont {Riedy}}, \bibinfo
  {author} {\bibfnamefont {D.~A.}\ \bibnamefont {Bader}}, \bibinfo {author}
  {\bibfnamefont {C.}~\bibnamefont {Corley}}, \bibinfo {author} {\bibfnamefont
  {R.}~\bibnamefont {Farber}},\ and\ \bibinfo {author} {\bibfnamefont {W.~N.}\
  \bibnamefont {Reynolds}},\ }\bibfield  {title} {\bibinfo {title} {Massive
  {Social} {Network} {Analysis}: {Mining} {Twitter} for {Social} {Good}},\ }in\
  \href {https://doi.org/10.1109/ICPP.2010.66} {\emph {\bibinfo {booktitle}
  {2010 39th {International} {Conference} on {Parallel} {Processing}}}}\
  (\bibinfo {year} {2010})\ pp.\ \bibinfo {pages} {583--593},\ \bibinfo {note}
  {iSSN: 2332-5690}\BibitemShut {NoStop}%
\bibitem [{\citenamefont {Traud}\ \emph {et~al.}(2012)\citenamefont {Traud},
  \citenamefont {Mucha},\ and\ \citenamefont {Porter}}]{traud_social_2012}%
  \BibitemOpen
  \bibfield  {author} {\bibinfo {author} {\bibfnamefont {A.~L.}\ \bibnamefont
  {Traud}}, \bibinfo {author} {\bibfnamefont {P.~J.}\ \bibnamefont {Mucha}},\
  and\ \bibinfo {author} {\bibfnamefont {M.~A.}\ \bibnamefont {Porter}},\
  }\bibfield  {title} {\bibinfo {title} {Social structure of {Facebook}
  networks},\ }\href {https://doi.org/10.1016/j.physa.2011.12.021} {\bibfield
  {journal} {\bibinfo  {journal} {Physica A: Statistical Mechanics and its
  Applications}\ }\textbf {\bibinfo {volume} {391}},\ \bibinfo {pages} {4165}
  (\bibinfo {year} {2012})}\BibitemShut {NoStop}%
\bibitem [{\citenamefont {Stam}\ and\ \citenamefont
  {Reijneveld}(2007)}]{stam_graph_2007}%
  \BibitemOpen
  \bibfield  {author} {\bibinfo {author} {\bibfnamefont {C.~J.}\ \bibnamefont
  {Stam}}\ and\ \bibinfo {author} {\bibfnamefont {J.~C.}\ \bibnamefont
  {Reijneveld}},\ }\bibfield  {title} {\bibinfo {title} {Graph theoretical
  analysis of complex networks in the brain},\ }\href
  {https://doi.org/10.1186/1753-4631-1-3} {\bibfield  {journal} {\bibinfo
  {journal} {Nonlinear Biomedical Physics}\ }\textbf {\bibinfo {volume} {1}},\
  \bibinfo {pages} {3} (\bibinfo {year} {2007})}\BibitemShut {NoStop}%
\bibitem [{\citenamefont {Pavlopoulos}\ \emph {et~al.}(2011)\citenamefont
  {Pavlopoulos}, \citenamefont {Secrier}, \citenamefont {Moschopoulos},
  \citenamefont {Soldatos}, \citenamefont {Kossida}, \citenamefont {Aerts},
  \citenamefont {Schneider},\ and\ \citenamefont
  {Bagos}}]{pavlopoulos_using_2011}%
  \BibitemOpen
  \bibfield  {author} {\bibinfo {author} {\bibfnamefont {G.~A.}\ \bibnamefont
  {Pavlopoulos}}, \bibinfo {author} {\bibfnamefont {M.}~\bibnamefont
  {Secrier}}, \bibinfo {author} {\bibfnamefont {C.~N.}\ \bibnamefont
  {Moschopoulos}}, \bibinfo {author} {\bibfnamefont {T.~G.}\ \bibnamefont
  {Soldatos}}, \bibinfo {author} {\bibfnamefont {S.}~\bibnamefont {Kossida}},
  \bibinfo {author} {\bibfnamefont {J.}~\bibnamefont {Aerts}}, \bibinfo
  {author} {\bibfnamefont {R.}~\bibnamefont {Schneider}},\ and\ \bibinfo
  {author} {\bibfnamefont {P.~G.}\ \bibnamefont {Bagos}},\ }\bibfield  {title}
  {\bibinfo {title} {Using graph theory to analyze biological networks},\
  }\href {https://doi.org/10.1186/1756-0381-4-10} {\bibfield  {journal}
  {\bibinfo  {journal} {BioData Mining}\ }\textbf {\bibinfo {volume} {4}},\
  \bibinfo {pages} {10} (\bibinfo {year} {2011})}\BibitemShut {NoStop}%
\bibitem [{\citenamefont {Farahani}\ \emph {et~al.}(2019)\citenamefont
  {Farahani}, \citenamefont {Karwowski},\ and\ \citenamefont
  {Lighthall}}]{farahani_application_2019}%
  \BibitemOpen
  \bibfield  {author} {\bibinfo {author} {\bibfnamefont {F.~V.}\ \bibnamefont
  {Farahani}}, \bibinfo {author} {\bibfnamefont {W.}~\bibnamefont
  {Karwowski}},\ and\ \bibinfo {author} {\bibfnamefont {N.~R.}\ \bibnamefont
  {Lighthall}},\ }\bibfield  {title} {\bibinfo {title} {Application of {Graph}
  {Theory} for {Identifying} {Connectivity} {Patterns} in {Human} {Brain}
  {Networks}: {A} {Systematic} {Review}},\ }\bibfield  {journal} {\bibinfo
  {journal} {Frontiers in Neuroscience}\ }\textbf {\bibinfo {volume} {13}},\
  \href {https://doi.org/10.3389/fnins.2019.00585} {10.3389/fnins.2019.00585}
  (\bibinfo {year} {2019}),\ \bibinfo {note} {publisher: Frontiers}\BibitemShut
  {NoStop}%
\bibitem [{\citenamefont {Monge}\ and\ \citenamefont
  {Contractor}(2003)}]{monge_theories_2003}%
  \BibitemOpen
  \bibfield  {author} {\bibinfo {author} {\bibfnamefont {P.~R.}\ \bibnamefont
  {Monge}}\ and\ \bibinfo {author} {\bibfnamefont {N.~S.}\ \bibnamefont
  {Contractor}},\ }\href
  {https://books.google.com/books?hl=de&lr=&id=5z3oPq8M5NwC&oi=fnd&pg=PR3&dq=communication+networks+graph+analysis&ots=H6akyppTnP&sig=-omwp5NkwWCxEKn-oZ0Xx8JmKRA}
  {\emph {\bibinfo {title} {Theories of communication networks}}}\ (\bibinfo
  {publisher} {Oxford University Press, USA},\ \bibinfo {year}
  {2003})\BibitemShut {NoStop}%
\bibitem [{\citenamefont {Hein}\ \emph {et~al.}(2006)\citenamefont {Hein},
  \citenamefont {Dür}, \citenamefont {Eisert}, \citenamefont {Raussendorf},
  \citenamefont {Van~den Nest},\ and\ \citenamefont
  {Briegel}}]{hein_entanglement_2006}%
  \BibitemOpen
  \bibfield  {author} {\bibinfo {author} {\bibfnamefont {M.}~\bibnamefont
  {Hein}}, \bibinfo {author} {\bibfnamefont {W.}~\bibnamefont {Dür}}, \bibinfo
  {author} {\bibfnamefont {J.}~\bibnamefont {Eisert}}, \bibinfo {author}
  {\bibfnamefont {R.}~\bibnamefont {Raussendorf}}, \bibinfo {author}
  {\bibfnamefont {M.}~\bibnamefont {Van~den Nest}},\ and\ \bibinfo {author}
  {\bibfnamefont {H.-J.}\ \bibnamefont {Briegel}},\ }\bibfield  {title}
  {\bibinfo {title} {Entanglement in graph states and its applications},\
  }\href {https://doi.org/10.3254/978-1-61499-018-5-115} {\bibfield  {journal}
  {\bibinfo  {journal} {Proceedings of the International School of Physics
  “Enrico Fermi”}\ }\textbf {\bibinfo {volume} {162}},\ \bibinfo {pages}
  {115} (\bibinfo {year} {2006})}\BibitemShut {NoStop}%
\bibitem [{\citenamefont {Coecke}\ and\ \citenamefont
  {Kissinger}(2010)}]{coecke_compositional_2010}%
  \BibitemOpen
  \bibfield  {author} {\bibinfo {author} {\bibfnamefont {B.}~\bibnamefont
  {Coecke}}\ and\ \bibinfo {author} {\bibfnamefont {A.}~\bibnamefont
  {Kissinger}},\ }\bibfield  {title} {\bibinfo {title} {The {Compositional}
  {Structure} of {Multipartite} {Quantum} {Entanglement}},\ }in\ \href
  {https://doi.org/10.1007/978-3-642-14162-1_25} {\emph {\bibinfo {booktitle}
  {Automata, {Languages} and {Programming}}}},\ \bibinfo {editor} {edited by\
  \bibinfo {editor} {\bibfnamefont {S.}~\bibnamefont {Abramsky}}, \bibinfo
  {editor} {\bibfnamefont {C.}~\bibnamefont {Gavoille}}, \bibinfo {editor}
  {\bibfnamefont {C.}~\bibnamefont {Kirchner}}, \bibinfo {editor}
  {\bibfnamefont {F.}~\bibnamefont {Meyer auf~der Heide}},\ and\ \bibinfo
  {editor} {\bibfnamefont {P.~G.}\ \bibnamefont {Spirakis}}}\ (\bibinfo
  {publisher} {Springer},\ \bibinfo {address} {Berlin, Heidelberg},\ \bibinfo
  {year} {2010})\ pp.\ \bibinfo {pages} {297--308}\BibitemShut {NoStop}%
\bibitem [{\citenamefont {de~Beaudrap}\ and\ \citenamefont
  {Roetteler}(2014)}]{de_beaudrap_quantum_2014}%
  \BibitemOpen
  \bibfield  {author} {\bibinfo {author} {\bibfnamefont {N.}~\bibnamefont
  {de~Beaudrap}}\ and\ \bibinfo {author} {\bibfnamefont {M.}~\bibnamefont
  {Roetteler}},\ }\bibfield  {title} {\bibinfo {title} {Quantum {Linear}
  {Network} {Coding} as {One}-way {Quantum} {Computation}},\ }in\ \href
  {https://doi.org/10.4230/LIPIcs.TQC.2014.217} {\emph {\bibinfo {booktitle}
  {9th {Conference} on the {Theory} of {Quantum} {Computation}, {Communication}
  and {Cryptography} ({TQC} 2014)}}}\ (\bibinfo  {publisher} {Schloss Dagstuhl
  – Leibniz-Zentrum für Informatik},\ \bibinfo {year} {2014})\ pp.\ \bibinfo
  {pages} {217--233}\BibitemShut {NoStop}%
\bibitem [{\citenamefont {Gnatenko}\ and\ \citenamefont
  {Susulovska}(2022)}]{gnatenko_geometric_2022}%
  \BibitemOpen
  \bibfield  {author} {\bibinfo {author} {\bibfnamefont {K.~P.}\ \bibnamefont
  {Gnatenko}}\ and\ \bibinfo {author} {\bibfnamefont {N.~A.}\ \bibnamefont
  {Susulovska}},\ }\bibfield  {title} {\bibinfo {title} {Geometric measure of
  entanglement of multi-qubit graph states and its detection on a quantum
  computer},\ }\href {https://doi.org/10.1209/0295-5075/ac419b} {\bibfield
  {journal} {\bibinfo  {journal} {Europhysics Letters}\ }\textbf {\bibinfo
  {volume} {136}},\ \bibinfo {pages} {40003} (\bibinfo {year} {2022})},\
  \bibinfo {note} {publisher: EDP Sciences, IOP Publishing and Società
  Italiana di Fisica}\BibitemShut {NoStop}%
\bibitem [{\citenamefont {Vesperini}\ and\ \citenamefont
  {Franzosi}(2024)}]{vesperini_entanglement_2024}%
  \BibitemOpen
  \bibfield  {author} {\bibinfo {author} {\bibfnamefont {A.}~\bibnamefont
  {Vesperini}}\ and\ \bibinfo {author} {\bibfnamefont {R.}~\bibnamefont
  {Franzosi}},\ }\bibfield  {title} {\bibinfo {title} {Entanglement, {Quantum}
  {Correlators}, and {Connectivity} in {Graph} {States}},\ }\href
  {https://doi.org/10.1002/qute.202300264} {\bibfield  {journal} {\bibinfo
  {journal} {Advanced Quantum Technologies}\ }\textbf {\bibinfo {volume} {7}},\
  \bibinfo {pages} {2300264} (\bibinfo {year} {2024})}\BibitemShut {NoStop}%
\bibitem [{\citenamefont {Gross}\ and\ \citenamefont
  {Yellen}(2004)}]{gross_handbook_2004}%
  \BibitemOpen
  \bibinfo {editor} {\bibfnamefont {J.~L.}\ \bibnamefont {Gross}}\ and\
  \bibinfo {editor} {\bibfnamefont {J.}~\bibnamefont {Yellen}},\ eds.,\
  \href@noop {} {\emph {\bibinfo {title} {Handbook of graph theory}}},\
  Discrete mathematics and its applications\ (\bibinfo  {publisher} {CRC
  Press},\ \bibinfo {address} {Boca Raton, Fla.},\ \bibinfo {year}
  {2004})\BibitemShut {NoStop}%
\bibitem [{\citenamefont {Ball}\ \emph {et~al.}(1995)\citenamefont {Ball},
  \citenamefont {Colbourn},\ and\ \citenamefont {Provan}}]{ball_network_1995}%
  \BibitemOpen
  \bibfield  {author} {\bibinfo {author} {\bibfnamefont {M.~O.}\ \bibnamefont
  {Ball}}, \bibinfo {author} {\bibfnamefont {C.~J.}\ \bibnamefont {Colbourn}},\
  and\ \bibinfo {author} {\bibfnamefont {J.~S.}\ \bibnamefont {Provan}},\
  }\bibfield  {title} {\bibinfo {title} {Network reliability},\ }in\ \href
  {https://doi.org/10.1016/S0927-0507(05)80128-8} {\emph {\bibinfo {booktitle}
  {Handbooks in {Operations} {Research} and {Management} {Science}}}},\
  \bibinfo {series} {Network {Models}}, Vol.~\bibinfo {volume} {7}\ (\bibinfo
  {publisher} {Elsevier},\ \bibinfo {year} {1995})\ pp.\ \bibinfo {pages}
  {673--762}\BibitemShut {NoStop}%
\bibitem [{\citenamefont {Schaeffer}(2007)}]{schaeffer_graph_2007}%
  \BibitemOpen
  \bibfield  {author} {\bibinfo {author} {\bibfnamefont {S.~E.}\ \bibnamefont
  {Schaeffer}},\ }\bibfield  {title} {\bibinfo {title} {Graph clustering},\
  }\href {https://doi.org/10.1016/j.cosrev.2007.05.001} {\bibfield  {journal}
  {\bibinfo  {journal} {Computer Science Review}\ }\textbf {\bibinfo {volume}
  {1}},\ \bibinfo {pages} {27} (\bibinfo {year} {2007})}\BibitemShut {NoStop}%
\bibitem [{\citenamefont {Erciyes}(2013)}]{erciyes_distributed_2013}%
  \BibitemOpen
  \bibfield  {author} {\bibinfo {author} {\bibfnamefont {K.}~\bibnamefont
  {Erciyes}},\ }\href@noop {} {\emph {\bibinfo {title} {Distributed {Graph}
  {Algorithms} for {Computer} {Networks}}}}\ (\bibinfo  {publisher} {Springer
  Science \& Business Media},\ \bibinfo {year} {2013})\BibitemShut {NoStop}%
\bibitem [{\citenamefont {Shiloach}\ and\ \citenamefont
  {Vishkin}(1982)}]{shiloach_ologn_1982}%
  \BibitemOpen
  \bibfield  {author} {\bibinfo {author} {\bibfnamefont {Y.}~\bibnamefont
  {Shiloach}}\ and\ \bibinfo {author} {\bibfnamefont {U.}~\bibnamefont
  {Vishkin}},\ }\bibfield  {title} {\bibinfo {title} {An {O}(logn) parallel
  connectivity algorithm},\ }\href
  {https://doi.org/10.1016/0196-6774(82)90008-6} {\bibfield  {journal}
  {\bibinfo  {journal} {Journal of Algorithms}\ }\textbf {\bibinfo {volume}
  {3}},\ \bibinfo {pages} {57} (\bibinfo {year} {1982})}\BibitemShut {NoStop}%
\bibitem [{\citenamefont {Dürr}\ \emph {et~al.}(2006)\citenamefont {Dürr},
  \citenamefont {Heiligman}, \citenamefont {Høyer},\ and\ \citenamefont
  {Mhalla}}]{durr_quantum_2006}%
  \BibitemOpen
  \bibfield  {author} {\bibinfo {author} {\bibfnamefont {C.}~\bibnamefont
  {Dürr}}, \bibinfo {author} {\bibfnamefont {M.}~\bibnamefont {Heiligman}},
  \bibinfo {author} {\bibfnamefont {P.}~\bibnamefont {Høyer}},\ and\ \bibinfo
  {author} {\bibfnamefont {M.}~\bibnamefont {Mhalla}},\ }\bibfield  {title}
  {\bibinfo {title} {Quantum {Query} {Complexity} of {Some} {Graph}
  {Problems}},\ }\href {https://doi.org/10.1137/050644719} {\bibfield
  {journal} {\bibinfo  {journal} {SIAM Journal on Computing}\ }\textbf
  {\bibinfo {volume} {35}},\ \bibinfo {pages} {1310} (\bibinfo {year}
  {2006})},\ \bibinfo {note} {publisher: Society for Industrial and Applied
  Mathematics}\BibitemShut {NoStop}%
\bibitem [{\citenamefont {Grover}(1996)}]{grover_fast_1996}%
  \BibitemOpen
  \bibfield  {author} {\bibinfo {author} {\bibfnamefont {L.~K.}\ \bibnamefont
  {Grover}},\ }\bibfield  {title} {\bibinfo {title} {A fast quantum mechanical
  algorithm for database search},\ }in\ \href
  {https://doi.org/10.1145/237814.237866} {\emph {\bibinfo {booktitle}
  {Proceedings of the twenty-eighth annual {ACM} symposium on {Theory} of
  {Computing}}}},\ \bibinfo {series and number} {{STOC} '96}\ (\bibinfo
  {publisher} {Association for Computing Machinery},\ \bibinfo {address} {New
  York, NY, USA},\ \bibinfo {year} {1996})\ pp.\ \bibinfo {pages}
  {212--219}\BibitemShut {NoStop}%
\bibitem [{\citenamefont {Coecke}\ and\ \citenamefont
  {Duncan}(2008)}]{coecke_interacting_2008}%
  \BibitemOpen
  \bibfield  {author} {\bibinfo {author} {\bibfnamefont {B.}~\bibnamefont
  {Coecke}}\ and\ \bibinfo {author} {\bibfnamefont {R.}~\bibnamefont
  {Duncan}},\ }\bibfield  {title} {\bibinfo {title} {Interacting {Quantum}
  {Observables}},\ }in\ \href {https://doi.org/10.1007/978-3-540-70583-3_25}
  {\emph {\bibinfo {booktitle} {Automata, {Languages} and {Programming}}}},\
  \bibinfo {editor} {edited by\ \bibinfo {editor} {\bibfnamefont
  {L.}~\bibnamefont {Aceto}}, \bibinfo {editor} {\bibfnamefont
  {I.}~\bibnamefont {Damgård}}, \bibinfo {editor} {\bibfnamefont {L.~A.}\
  \bibnamefont {Goldberg}}, \bibinfo {editor} {\bibfnamefont {M.~M.}\
  \bibnamefont {Halldórsson}}, \bibinfo {editor} {\bibfnamefont
  {A.}~\bibnamefont {Ingólfsdóttir}},\ and\ \bibinfo {editor} {\bibfnamefont
  {I.}~\bibnamefont {Walukiewicz}}}\ (\bibinfo  {publisher} {Springer},\
  \bibinfo {address} {Berlin, Heidelberg},\ \bibinfo {year} {2008})\ pp.\
  \bibinfo {pages} {298--310}\BibitemShut {NoStop}%
\bibitem [{\citenamefont {Coecke}\ and\ \citenamefont
  {Kissinger}(2017)}]{coecke_picturing_2017}%
  \BibitemOpen
  \bibfield  {author} {\bibinfo {author} {\bibfnamefont {B.}~\bibnamefont
  {Coecke}}\ and\ \bibinfo {author} {\bibfnamefont {A.}~\bibnamefont
  {Kissinger}},\ }\href {https://doi.org/10.1017/9781316219317} {\emph
  {\bibinfo {title} {Picturing {Quantum} {Processes}: {A} {First} {Course} in
  {Quantum} {Theory} and {Diagrammatic} {Reasoning}}}}\ (\bibinfo  {publisher}
  {Cambridge University Press},\ \bibinfo {address} {Cambridge},\ \bibinfo
  {year} {2017})\BibitemShut {NoStop}%
\bibitem [{\citenamefont {van~de
  Wetering}(2020)}]{van_de_wetering_zx-calculus_2020}%
  \BibitemOpen
  \bibfield  {author} {\bibinfo {author} {\bibfnamefont {J.}~\bibnamefont
  {van~de Wetering}},\ }\href {http://arxiv.org/abs/2012.13966} {\bibinfo
  {title} {{ZX}-calculus for the working quantum computer scientist}} (\bibinfo
  {year} {2020}),\ \bibinfo {note} {arXiv:2012.13966 [quant-ph]}\BibitemShut
  {NoStop}%
\bibitem [{\citenamefont {Coecke}\ and\ \citenamefont
  {Duncan}(2011)}]{coecke_interacting_2011}%
  \BibitemOpen
  \bibfield  {author} {\bibinfo {author} {\bibfnamefont {B.}~\bibnamefont
  {Coecke}}\ and\ \bibinfo {author} {\bibfnamefont {R.}~\bibnamefont
  {Duncan}},\ }\bibfield  {title} {\bibinfo {title} {Interacting quantum
  observables: categorical algebra and diagrammatics},\ }\href@noop {}
  {\bibfield  {journal} {\bibinfo  {journal} {New Journal of Physics}\ }\textbf
  {\bibinfo {volume} {13}},\ \bibinfo {pages} {043016} (\bibinfo {year}
  {2011})},\ \bibinfo {note} {publisher: IOP Publishing}\BibitemShut {NoStop}%
\bibitem [{\citenamefont {Vilmart}(2019)}]{vilmart-near-optimal-2018}%
  \BibitemOpen
  \bibfield  {author} {\bibinfo {author} {\bibfnamefont {R.}~\bibnamefont
  {Vilmart}},\ }\bibfield  {title} {\bibinfo {title} {A near-minimal
  axiomatisation of zx-calculus for pure qubit quantum mechanics},\ }in\ \href
  {https://doi.org/10.1109/LICS.2019.8785765} {\emph {\bibinfo {booktitle}
  {2019 34th Annual ACM/IEEE Symposium on Logic in Computer Science (LICS)}}}\
  (\bibinfo {year} {2019})\ pp.\ \bibinfo {pages} {1--10}\BibitemShut {NoStop}%
\bibitem [{\citenamefont {Wang}\ and\ \citenamefont
  {Yeung}(2023)}]{wang_representing_2023}%
  \BibitemOpen
  \bibfield  {author} {\bibinfo {author} {\bibfnamefont {Q.}~\bibnamefont
  {Wang}}\ and\ \bibinfo {author} {\bibfnamefont {R.}~\bibnamefont {Yeung}},\
  }\href {https://doi.org/10.48550/arXiv.2110.06898} {\bibinfo {title}
  {Representing and {Implementing} {Matrices} {Using} {Algebraic}
  {ZX}-calculus}} (\bibinfo {year} {2023}),\ \bibinfo {note}
  {arXiv:2110.06898}\BibitemShut {NoStop}%
\bibitem [{\citenamefont {Terashima}\ and\ \citenamefont
  {Ueda}(2005)}]{terashima_nonunitary_2005}%
  \BibitemOpen
  \bibfield  {author} {\bibinfo {author} {\bibfnamefont {H.}~\bibnamefont
  {Terashima}}\ and\ \bibinfo {author} {\bibfnamefont {M.}~\bibnamefont
  {Ueda}},\ }\bibfield  {title} {\bibinfo {title} {Nonunitary quantum
  circuit},\ }\href {https://doi.org/10.1142/S0219749905001456} {\bibfield
  {journal} {\bibinfo  {journal} {International Journal of Quantum
  Information}\ }\textbf {\bibinfo {volume} {03}},\ \bibinfo {pages} {633}
  (\bibinfo {year} {2005})},\ \bibinfo {note} {publisher: World Scientific
  Publishing Co.}\BibitemShut {Stop}%
\bibitem [{\citenamefont {Brassard}\ \emph {et~al.}(2002)\citenamefont
  {Brassard}, \citenamefont {Høyer}, \citenamefont {Mosca},\ and\
  \citenamefont {Tapp}}]{brassard_quantum_2002}%
  \BibitemOpen
  \bibfield  {author} {\bibinfo {author} {\bibfnamefont {G.}~\bibnamefont
  {Brassard}}, \bibinfo {author} {\bibfnamefont {P.}~\bibnamefont {Høyer}},
  \bibinfo {author} {\bibfnamefont {M.}~\bibnamefont {Mosca}},\ and\ \bibinfo
  {author} {\bibfnamefont {A.}~\bibnamefont {Tapp}},\ }\bibfield  {title}
  {\bibinfo {title} {Quantum amplitude amplification and estimation},\ }\href
  {https://books.google.com/books?hl=de&lr=&id=o2QbCAAAQBAJ&oi=fnd&pg=PA53&dq=G.+Brassard,+P.+H%C3%B8yer,+M.+Mosca+and+A.+Tapp.+Quantum+amplitude+amplification+and+estimation.+In+Quantum+Computation+and+Quantum+Information:+A+Millennium+Volume,+AMS+Contemporary+Mathematics+Series.&ots=0AW_5gWu8r&sig=lH_LiDSC7ERxNabaR55UGgfX0G4}
  {\bibfield  {journal} {\bibinfo  {journal} {Contemporary Mathematics}\
  }\textbf {\bibinfo {volume} {305}},\ \bibinfo {pages} {53} (\bibinfo {year}
  {2002})},\ \bibinfo {note} {publisher: Providence, RI; American Mathematical
  Society; 1999}\BibitemShut {NoStop}%
\bibitem [{\citenamefont {Knill}(2002)}]{knill_quantum_2002}%
  \BibitemOpen
  \bibfield  {author} {\bibinfo {author} {\bibfnamefont {E.}~\bibnamefont
  {Knill}},\ }\bibfield  {title} {\bibinfo {title} {Quantum gates using linear
  optics and postselection},\ }\bibfield  {journal} {\bibinfo  {journal}
  {Physical Review A}\ }\textbf {\bibinfo {volume} {66}},\ \href
  {https://doi.org/10.1103/PhysRevA.66.052306} {10.1103/PhysRevA.66.052306}
  (\bibinfo {year} {2002})\BibitemShut {NoStop}%
\bibitem [{\citenamefont {Christandl}(2009)}]{christandl_postselection_2009}%
  \BibitemOpen
  \bibfield  {author} {\bibinfo {author} {\bibfnamefont {M.}~\bibnamefont
  {Christandl}},\ }\bibfield  {title} {\bibinfo {title} {Postselection
  {Technique} for {Quantum} {Channels} with {Applications} to {Quantum}
  {Cryptography}},\ }\bibfield  {journal} {\bibinfo  {journal} {Physical Review
  Letters}\ }\textbf {\bibinfo {volume} {102}},\ \href
  {https://doi.org/10.1103/PhysRevLett.102.020504}
  {10.1103/PhysRevLett.102.020504} (\bibinfo {year} {2009})\BibitemShut
  {NoStop}%
\bibitem [{\citenamefont {Renner}(2010)}]{renner_simplifying_2010}%
  \BibitemOpen
  \bibfield  {author} {\bibinfo {author} {\bibfnamefont {R.}~\bibnamefont
  {Renner}},\ }\bibfield  {title} {\bibinfo {title} {Simplifying
  information-theoretic arguments by post-selection},\ }in\ \href
  {https://doi.org/10.3233/978-1-60750-547-1-66} {\emph {\bibinfo {booktitle}
  {Quantum {Cryptography} and {Computing}}}}\ (\bibinfo  {publisher} {IOS
  Press},\ \bibinfo {year} {2010})\ pp.\ \bibinfo {pages} {66--75}\BibitemShut
  {NoStop}%
\bibitem [{\citenamefont {Wright}(2021)}]{wright_automatic_2021}%
  \BibitemOpen
  \bibfield  {author} {\bibinfo {author} {\bibfnamefont {L.}~\bibnamefont
  {Wright}},\ }\bibfield  {title} {\bibinfo {title} {Automatic post-selection
  by ancillae thermalization},\ }\bibfield  {journal} {\bibinfo  {journal}
  {Physical Review Research}\ }\textbf {\bibinfo {volume} {3}},\ \href
  {https://doi.org/10.1103/PhysRevResearch.3.033151}
  {10.1103/PhysRevResearch.3.033151} (\bibinfo {year} {2021})\BibitemShut
  {NoStop}%
\bibitem [{\citenamefont {Poulin}(2005)}]{poulin_stabilizer_2005}%
  \BibitemOpen
  \bibfield  {author} {\bibinfo {author} {\bibfnamefont {D.}~\bibnamefont
  {Poulin}},\ }\bibfield  {title} {\bibinfo {title} {Stabilizer {Formalism} for
  {Operator} {Quantum} {Error} {Correction}},\ }\bibfield  {journal} {\bibinfo
  {journal} {Physical Review Letters}\ }\textbf {\bibinfo {volume} {95}},\
  \href {https://doi.org/10.1103/PhysRevLett.95.230504}
  {10.1103/PhysRevLett.95.230504} (\bibinfo {year} {2005})\BibitemShut
  {NoStop}%
\bibitem [{\citenamefont {Abrams}\ and\ \citenamefont
  {Lloyd}(1998)}]{abrams_nonlinear_1998}%
  \BibitemOpen
  \bibfield  {author} {\bibinfo {author} {\bibfnamefont {D.~S.}\ \bibnamefont
  {Abrams}}\ and\ \bibinfo {author} {\bibfnamefont {S.}~\bibnamefont {Lloyd}},\
  }\bibfield  {title} {\bibinfo {title} {Nonlinear {Quantum} {Mechanics}
  {Implies} {Polynomial}-{Time} {Solution} for
  \${\textbackslash}mathit\{{NP}\}\$-{Complete} and \#
  \${\textbackslash}mathit\{{P}\}\$ {Problems}},\ }\href
  {https://doi.org/10.1103/PhysRevLett.81.3992} {\bibfield  {journal} {\bibinfo
   {journal} {Physical Review Letters}\ }\textbf {\bibinfo {volume} {81}},\
  \bibinfo {pages} {3992} (\bibinfo {year} {1998})},\ \bibinfo {note}
  {publisher: American Physical Society}\BibitemShut {NoStop}%
\bibitem [{\citenamefont {Song}\ \emph {et~al.}(2019)\citenamefont {Song},
  \citenamefont {Xu}, \citenamefont {Li}, \citenamefont {Zhang}, \citenamefont
  {Zhang}, \citenamefont {Liu}, \citenamefont {Guo}, \citenamefont {Wang},
  \citenamefont {Ren}, \citenamefont {Hao}, \citenamefont {Feng}, \citenamefont
  {Fan}, \citenamefont {Zheng}, \citenamefont {Wang}, \citenamefont {Wang},\
  and\ \citenamefont {Zhu}}]{song_generation_2019}%
  \BibitemOpen
  \bibfield  {author} {\bibinfo {author} {\bibfnamefont {C.}~\bibnamefont
  {Song}}, \bibinfo {author} {\bibfnamefont {K.}~\bibnamefont {Xu}}, \bibinfo
  {author} {\bibfnamefont {H.}~\bibnamefont {Li}}, \bibinfo {author}
  {\bibfnamefont {Y.-R.}\ \bibnamefont {Zhang}}, \bibinfo {author}
  {\bibfnamefont {X.}~\bibnamefont {Zhang}}, \bibinfo {author} {\bibfnamefont
  {W.}~\bibnamefont {Liu}}, \bibinfo {author} {\bibfnamefont {Q.}~\bibnamefont
  {Guo}}, \bibinfo {author} {\bibfnamefont {Z.}~\bibnamefont {Wang}}, \bibinfo
  {author} {\bibfnamefont {W.}~\bibnamefont {Ren}}, \bibinfo {author}
  {\bibfnamefont {J.}~\bibnamefont {Hao}}, \bibinfo {author} {\bibfnamefont
  {H.}~\bibnamefont {Feng}}, \bibinfo {author} {\bibfnamefont {H.}~\bibnamefont
  {Fan}}, \bibinfo {author} {\bibfnamefont {D.}~\bibnamefont {Zheng}}, \bibinfo
  {author} {\bibfnamefont {D.-W.}\ \bibnamefont {Wang}}, \bibinfo {author}
  {\bibfnamefont {H.}~\bibnamefont {Wang}},\ and\ \bibinfo {author}
  {\bibfnamefont {S.-Y.}\ \bibnamefont {Zhu}},\ }\bibfield  {title} {\bibinfo
  {title} {Generation of multicomponent atomic {Schrödinger} cat states of up
  to 20 qubits},\ }\href {https://doi.org/10.1126/science.aay0600} {\bibfield
  {journal} {\bibinfo  {journal} {Science}\ }\textbf {\bibinfo {volume}
  {365}},\ \bibinfo {pages} {574} (\bibinfo {year} {2019})},\ \bibinfo {note}
  {publisher: American Association for the Advancement of Science}\BibitemShut
  {NoStop}%
\bibitem [{\citenamefont {Mooney}\ \emph
  {et~al.}(2021{\natexlab{a}})\citenamefont {Mooney}, \citenamefont {White},
  \citenamefont {Hill},\ and\ \citenamefont
  {Hollenberg}}]{mooney_generation_2021}%
  \BibitemOpen
  \bibfield  {author} {\bibinfo {author} {\bibfnamefont {G.~J.}\ \bibnamefont
  {Mooney}}, \bibinfo {author} {\bibfnamefont {G.~A.~L.}\ \bibnamefont
  {White}}, \bibinfo {author} {\bibfnamefont {C.~D.}\ \bibnamefont {Hill}},\
  and\ \bibinfo {author} {\bibfnamefont {L.~C.~L.}\ \bibnamefont
  {Hollenberg}},\ }\bibfield  {title} {\bibinfo {title} {Generation and
  verification of 27-qubit {Greenberger}-{Horne}-{Zeilinger} states in a
  superconducting quantum computer},\ }\href
  {https://doi.org/10.1088/2399-6528/ac1df7} {\bibfield  {journal} {\bibinfo
  {journal} {Journal of Physics Communications}\ }\textbf {\bibinfo {volume}
  {5}},\ \bibinfo {pages} {095004} (\bibinfo {year} {2021}{\natexlab{a}})},\
  \bibinfo {note} {publisher: IOP Publishing}\BibitemShut {NoStop}%
\bibitem [{\citenamefont {Mooney}\ \emph
  {et~al.}(2021{\natexlab{b}})\citenamefont {Mooney}, \citenamefont {White},
  \citenamefont {Hill},\ and\ \citenamefont
  {Hollenberg}}]{mooney_whole-device_2021}%
  \BibitemOpen
  \bibfield  {author} {\bibinfo {author} {\bibfnamefont {G.~J.}\ \bibnamefont
  {Mooney}}, \bibinfo {author} {\bibfnamefont {G.~A.~L.}\ \bibnamefont
  {White}}, \bibinfo {author} {\bibfnamefont {C.~D.}\ \bibnamefont {Hill}},\
  and\ \bibinfo {author} {\bibfnamefont {L.~C.~L.}\ \bibnamefont
  {Hollenberg}},\ }\bibfield  {title} {\bibinfo {title} {Whole-{Device}
  {Entanglement} in a 65-{Qubit} {Superconducting} {Quantum} {Computer}},\
  }\href {https://doi.org/10.1002/qute.202100061} {\bibfield  {journal}
  {\bibinfo  {journal} {Advanced Quantum Technologies}\ }\textbf {\bibinfo
  {volume} {4}},\ \bibinfo {pages} {2100061} (\bibinfo {year}
  {2021}{\natexlab{b}})}\BibitemShut {NoStop}%
\bibitem [{\citenamefont {Zhao}\ \emph {et~al.}(2021)\citenamefont {Zhao},
  \citenamefont {Zhang}, \citenamefont {Chen}, \citenamefont {Wang},\ and\
  \citenamefont {Hu}}]{zhao_creation_2021}%
  \BibitemOpen
  \bibfield  {author} {\bibinfo {author} {\bibfnamefont {Y.}~\bibnamefont
  {Zhao}}, \bibinfo {author} {\bibfnamefont {R.}~\bibnamefont {Zhang}},
  \bibinfo {author} {\bibfnamefont {W.}~\bibnamefont {Chen}}, \bibinfo {author}
  {\bibfnamefont {X.-B.}\ \bibnamefont {Wang}},\ and\ \bibinfo {author}
  {\bibfnamefont {J.}~\bibnamefont {Hu}},\ }\bibfield  {title} {\bibinfo
  {title} {Creation of {Greenberger}-{Horne}-{Zeilinger} states with thousands
  of atoms by entanglement amplification},\ }\href
  {https://doi.org/10.1038/s41534-021-00364-8} {\bibfield  {journal} {\bibinfo
  {journal} {npj Quantum Information}\ }\textbf {\bibinfo {volume} {7}},\
  \bibinfo {pages} {1} (\bibinfo {year} {2021})},\ \bibinfo {note} {publisher:
  Nature Publishing Group}\BibitemShut {NoStop}%
\bibitem [{\citenamefont {Briegel}\ \emph {et~al.}(2009)\citenamefont
  {Briegel}, \citenamefont {Browne}, \citenamefont {Dür}, \citenamefont
  {Raussendorf},\ and\ \citenamefont {Van~den
  Nest}}]{briegel_measurement-based_2009}%
  \BibitemOpen
  \bibfield  {author} {\bibinfo {author} {\bibfnamefont {H.~J.}\ \bibnamefont
  {Briegel}}, \bibinfo {author} {\bibfnamefont {D.~E.}\ \bibnamefont {Browne}},
  \bibinfo {author} {\bibfnamefont {W.}~\bibnamefont {Dür}}, \bibinfo {author}
  {\bibfnamefont {R.}~\bibnamefont {Raussendorf}},\ and\ \bibinfo {author}
  {\bibfnamefont {M.}~\bibnamefont {Van~den Nest}},\ }\bibfield  {title}
  {\bibinfo {title} {Measurement-based quantum computation},\ }\href
  {https://doi.org/10.1038/nphys1157} {\bibfield  {journal} {\bibinfo
  {journal} {Nature Physics}\ }\textbf {\bibinfo {volume} {5}},\ \bibinfo
  {pages} {19} (\bibinfo {year} {2009})},\ \bibinfo {note} {publisher: Nature
  Publishing Group}\BibitemShut {NoStop}%
\bibitem [{\citenamefont {Kissinger}\ and\ \citenamefont
  {Wetering}(2019)}]{kissinger_universal_2019}%
  \BibitemOpen
  \bibfield  {author} {\bibinfo {author} {\bibfnamefont {A.}~\bibnamefont
  {Kissinger}}\ and\ \bibinfo {author} {\bibfnamefont {J.~v.~d.}\ \bibnamefont
  {Wetering}},\ }\bibfield  {title} {\bibinfo {title} {Universal {MBQC} with
  generalised parity-phase interactions and {Pauli} measurements},\ }\href
  {https://doi.org/10.22331/q-2019-04-26-134} {\bibfield  {journal} {\bibinfo
  {journal} {Quantum}\ }\textbf {\bibinfo {volume} {3}},\ \bibinfo {pages}
  {134} (\bibinfo {year} {2019})},\ \bibinfo {note} {publisher: Verein zur
  Förderung des Open Access Publizierens in den
  Quantenwissenschaften}\BibitemShut {NoStop}%
\bibitem [{\citenamefont {Duncan}\ and\ \citenamefont
  {Perdrix}(2010)}]{duncan_rewriting_2010}%
  \BibitemOpen
  \bibfield  {author} {\bibinfo {author} {\bibfnamefont {R.}~\bibnamefont
  {Duncan}}\ and\ \bibinfo {author} {\bibfnamefont {S.}~\bibnamefont
  {Perdrix}},\ }\bibfield  {title} {\bibinfo {title} {Rewriting
  {Measurement}-{Based} {Quantum} {Computations} with {Generalised} {Flow}},\
  }in\ \href {https://doi.org/10.1007/978-3-642-14162-1_24} {\emph {\bibinfo
  {booktitle} {Automata, {Languages} and {Programming}}}},\ \bibinfo {editor}
  {edited by\ \bibinfo {editor} {\bibfnamefont {S.}~\bibnamefont {Abramsky}},
  \bibinfo {editor} {\bibfnamefont {C.}~\bibnamefont {Gavoille}}, \bibinfo
  {editor} {\bibfnamefont {C.}~\bibnamefont {Kirchner}}, \bibinfo {editor}
  {\bibfnamefont {F.}~\bibnamefont {Meyer auf~der Heide}},\ and\ \bibinfo
  {editor} {\bibfnamefont {P.~G.}\ \bibnamefont {Spirakis}}}\ (\bibinfo
  {publisher} {Springer},\ \bibinfo {address} {Berlin, Heidelberg},\ \bibinfo
  {year} {2010})\ pp.\ \bibinfo {pages} {285--296}\BibitemShut {NoStop}%
\bibitem [{\citenamefont {Beckey}(2023)}]{beckey_multipartite_2023}%
  \BibitemOpen
  \bibfield  {author} {\bibinfo {author} {\bibfnamefont {J.~L.}\ \bibnamefont
  {Beckey}},\ }\bibfield  {title} {\bibinfo {title} {Multipartite entanglement
  measures via {Bell}-basis measurements},\ }\bibfield  {journal} {\bibinfo
  {journal} {Physical Review A}\ }\textbf {\bibinfo {volume} {107}},\ \href
  {https://doi.org/10.1103/PhysRevA.107.062425} {10.1103/PhysRevA.107.062425}
  (\bibinfo {year} {2023})\BibitemShut {NoStop}%
\bibitem [{\citenamefont {Bezanson}\ \emph {et~al.}(2017)\citenamefont
  {Bezanson}, \citenamefont {Edelman}, \citenamefont {Karpinski},\ and\
  \citenamefont {Shah}}]{bezanson2017julia}%
  \BibitemOpen
  \bibfield  {author} {\bibinfo {author} {\bibfnamefont {J.}~\bibnamefont
  {Bezanson}}, \bibinfo {author} {\bibfnamefont {A.}~\bibnamefont {Edelman}},
  \bibinfo {author} {\bibfnamefont {S.}~\bibnamefont {Karpinski}},\ and\
  \bibinfo {author} {\bibfnamefont {V.~B.}\ \bibnamefont {Shah}},\ }\bibfield
  {title} {\bibinfo {title} {Julia: A fresh approach to numerical computing},\
  }\href {https://doi.org/10.1137/141000671} {\bibfield  {journal} {\bibinfo
  {journal} {SIAM review}\ }\textbf {\bibinfo {volume} {59}},\ \bibinfo {pages}
  {65} (\bibinfo {year} {2017})}\BibitemShut {NoStop}%
\bibitem [{\citenamefont {Luo}\ \emph {et~al.}(2020)\citenamefont {Luo},
  \citenamefont {Liu}, \citenamefont {Zhang},\ and\ \citenamefont
  {Wang}}]{luo_yaojl_2020}%
  \BibitemOpen
  \bibfield  {author} {\bibinfo {author} {\bibfnamefont {X.-Z.}\ \bibnamefont
  {Luo}}, \bibinfo {author} {\bibfnamefont {J.-G.}\ \bibnamefont {Liu}},
  \bibinfo {author} {\bibfnamefont {P.}~\bibnamefont {Zhang}},\ and\ \bibinfo
  {author} {\bibfnamefont {L.}~\bibnamefont {Wang}},\ }\bibfield  {title}
  {\bibinfo {title} {Yao.jl: {Extensible}, {Efficient} {Framework} for
  {Quantum} {Algorithm} {Design}},\ }\href
  {https://doi.org/10.22331/q-2020-10-11-341} {\bibfield  {journal} {\bibinfo
  {journal} {Quantum}\ }\textbf {\bibinfo {volume} {4}},\ \bibinfo {pages}
  {341} (\bibinfo {year} {2020})},\ \bibinfo {note} {arXiv:
  1912.10877}\BibitemShut {NoStop}%
\bibitem [{\citenamefont {Erdős}\ and\ \citenamefont
  {Rényi}(1959)}]{erdos_random_1959}%
  \BibitemOpen
  \bibfield  {author} {\bibinfo {author} {\bibfnamefont {P.}~\bibnamefont
  {Erdős}}\ and\ \bibinfo {author} {\bibfnamefont {A.}~\bibnamefont
  {Rényi}},\ }\bibfield  {title} {\bibinfo {title} {On random graphs. {I}.},\
  }\href {https://doi.org/10.5486/PMD.1959.6.3-4.12} {\bibfield  {journal}
  {\bibinfo  {journal} {Publicationes Mathematicae Debrecen}\ }\textbf
  {\bibinfo {volume} {6}},\ \bibinfo {pages} {290} (\bibinfo {year}
  {1959})}\BibitemShut {NoStop}%
\bibitem [{\citenamefont {Mertens}(2005)}]{mertens_easiest_2005}%
  \BibitemOpen
  \bibfield  {author} {\bibinfo {author} {\bibfnamefont {S.}~\bibnamefont
  {Mertens}},\ }\bibfield  {title} {\bibinfo {title} {The {Easiest} {Hard}
  {Problem}: {Number} {Partitioning}},\ }in\ \href
  {https://doi.org/10.1093/oso/9780195177374.003.0012} {\emph {\bibinfo
  {booktitle} {Computational {Complexity} and {Statistical} {Physics}}}}\
  (\bibinfo  {publisher} {Oxford University Press},\ \bibinfo {year}
  {2005})\BibitemShut {NoStop}%
\end{thebibliography}
